\documentclass[lettersize,journal]{IEEEtran}
\usepackage{amsmath,amsfonts,amssymb}
\usepackage{graphicx}
\usepackage{subfig}
\usepackage{algorithmicx}
\usepackage{algorithm}
\usepackage{algpseudocode}
\usepackage{textcomp}
\usepackage{makecell}
\usepackage{multirow,multicol}
\usepackage[colorlinks]{hyperref}
\usepackage[table]{xcolor}
\usepackage{threeparttable}

\begin{document}

\title{Downlink OFDM-FAMA in 5G-NR Systems} 
\author{Hanjiang Hong,~\IEEEmembership{Member,~IEEE}, 
        Kai-Kit Wong,~\IEEEmembership{Fellow,~IEEE}, 
        Hao Xu,~\IEEEmembership{Member,~IEEE}, \\
        Yin Xu,~\IEEEmembership{Senior Member,~IEEE},
        Hyundong Shin,~\IEEEmembership{Fellow}, \textit{IEEE},  
        Ross Murch,~\IEEEmembership{Fellow}, \textit{IEEE}, \\
        Dazhi He,~\IEEEmembership{Senior Member,~IEEE}, and
        Wenjun Zhang,~\IEEEmembership{Fellow}, \textit{IEEE}
\vspace{-10mm}

\thanks{The work of K. K. Wong is supported by the Engineering and Physical Sciences Research Council (EPSRC) under Grant EP/W026813/1.}
\thanks{The work of H. Hong is supported by the Outstanding Doctoral Graduates Development Scholarship of Shanghai Jiao Tong University.}
\thanks{The work of R. Murch is supported by the Hong Kong Research Grants Council Area of Excellence grant AoE/E-601/22-R.}
\thanks{The work of Y. Xu and D. He is supported in part by National Natural Science Foundation of China under Grant 62271316 and 62422111.}
\thanks{H. Hong and K. K. Wong are with the Department of Electronic and Electrical Engineering, University College London, London, United Kingdom. K. K. Wong is also affiliated with the Department of Electronic Engineering, Kyung Hee University, Yongin-si, Gyeonggi-do 17104, Korea.}
\thanks{H. Xu is with the National Mobile Communications Research Laboratory, Southeast University, Nanjing 210096, China.}
\thanks{Y. Xu, D. He, and W. Zhang is with the Cooperative Medianet Innovation Center (CMIC), Shanghai Jiao Tong University, Shanghai, China.}
\thanks{H. Shin is with the Department of Electronic Engineering, Kyung Hee University, Yongin-si, Gyeonggi-do 17104, Korea.}
\thanks{R. Murch is with the Department of Electronic and Computer Engineering and Institute for Advanced Study (IAS), The Hong Kong University of Science and Technology, Clear Water Bay, Hong Kong SAR, China.}

\thanks{Corresponding author: Kai-Kit Wong.}
}
\maketitle

\begin{abstract}
Fluid antenna multiple access (FAMA), enabled by the fluid antenna system (FAS), offers a new and straightforward solution to massive connectivity. Previous results on FAMA were primarily based on narrowband channels. This paper studies the adoption of FAMA within the fifth-generation (5G) orthogonal frequency division multiplexing (OFDM) framework, referred to as OFDM-FAMA, and evaluate its performance in broadband multipath channels. We first design the OFDM-FAMA system, taking into account 5G channel coding and OFDM modulation. Then the system's achievable rate is analyzed, and an algorithm to approximate the FAS configuration at each user is proposed based on the rate. Extensive link-level simulation results reveal that OFDM-FAMA can significantly improve the multiplexing gain over the OFDM system with fixed-position antenna (FPA) users, especially when robust channel coding is applied and the number of radio-frequency (RF) chains at each user is small.
\end{abstract}

\begin{IEEEkeywords}
Fluid antenna multiple access (FAMA), fluid antenna system (FAS), OFDM, performance evaluation.
\end{IEEEkeywords}

\vspace{-3mm}
\section{Introduction}
\IEEEPARstart{E}{xtreme} massive connectivity is a cornerstone of massive machine-type communication (mMTC) in the fifth-generation (5G) communication system, which aims to provide connections to a very large number of Internet of Things (IoT) devices and facilitate sporadical transmission, e.g., \cite{mahmood2020six,chen2021massive,ngo2024ultradense}. These complicated scenarios inherent in mMTC often render the acquisition of channel state information (CSI) at the base station (BS) impractical, hampering the way in which massive connectivity could be achieved. Prominent techniques, such as massive multiple-input multiple-output (MIMO) \cite{wang2024extremely,pereira2022anOverview}, non-orthogonal multiple access (NOMA) \cite{wang2018aSurvey}, and rate-splitting multiple access (RSMA) \cite{mao2022rsma}, have shown promises but under these circumstances, they may not be feasible.

To improve scalability, we need a straightforward multiple access scheme that allows for massive spectrum sharing and requires less transmitter CSI. 
A promising approach to tackle this challenge is fluid antenna multiple access (FAMA) \cite{wong2022FAMA}. FAMA takes great advantage of the antenna position flexibility enabled by the fluid antenna system (FAS) technology to operate at the natural interference null for reception \cite{New2024aTutorial}.

The concept of FAS was proposed in 2020 by Wong {\em et al.} \cite{wong2020FAS,wong2021FAS}. Its implementation relies on flexible technologies that may be realized in the form of liquid-based antennas \cite{huang2021liquid, paracha2019liquid, shen2024design}, reconfigurable pixel-based antennas \cite{hoang2021computational,zhang2024pixel}, stepper motor-based antennas \cite{basbug2017design}, or flexible structures using metamaterials \cite{johnson2015sidelobe}. Recent studies in FAS have already investigated the performance in different channels \cite{Khammassi2023, Vega2023novel, Vega2023asimple, Alvim2023on, Psomas2023continuous, New2023fluid}. Channel estimation for FAS has also been studied recently recognizing the spatial correlation \cite{Skouroumounis2023fluid, xu2024channel,zhang2023successive}. Efforts have also been made to combine FAS with other technologies, e.g., MIMO \cite{new2023information}, integrated sensing and communication (ISAC) \cite{zhou2024fasisac}, full duplex \cite{Skouroumounis2023fasfd}, and etc.

For multiuser communications, FAMA was first introduced in 2022 \cite{wong2022FAMA}. It exploits the unique feature of FAS to access the spatial opportunity where the interference becomes weak. Depending on how fast the user updates the position (a.k.a.~port) of FAS, FAMA can be classified into {\em fast} FAMA ({\em f}-FAMA)  \cite{wong2022FAMA,wong2022fast} or {\em slow} FAMA ({\em s}-FAMA) \cite{wong2023sFAMA,Wong2024cuma,wong2024transmitter,espinosa2024anew,Xu2024revisiting,Waqar2023deep}. The {\em f}-FAMA scheme switches the antenna port on a per-symbol basis, where the data-dependant sum interference plus noise signal cancels. However, \textit{f}-FAMA could be impractical because of the complexity of instantaneously observing a large number of received signals. By contrast, {\em s}-FAMA is a more practical scheme because it only requires the FAS to switch the position once during each channel coherence time. Recently in \cite{hong2024coded, hong20245gcoded}, 5G New Radio (NR) coded modulation schemes have also been considered to improve FAMA. 

Nonetheless, state-of-art researches on FAS and FAMA have been based on narrowband channels with flat fading characteristics. But in real channels, broadband communications is more likely and the adoption of orthogonal frequency-division multiplexing (OFDM) is necessary to resolve the delay spread \cite{Kaakinen2021frequency}. Technically speaking, there is concern if FAS could still perform well in OFDM settings because a desirable port at a subcarrier is likely not as desirable for other subcarriers. It is important to find out if FAS is indeed useful for 5G NR.

Motivated by the above, this paper integrates FAMA within the OFDM system, resulting in the proposed OFDM-FAMA system. This framework enables the BS to effectively distribute the signals to multiple user terminals (UTs) without requiring CSI at the BS or the need of interference cancellation at the UTs. Our effort aligns with the 5G NR physical layer procedures, and proposes a port selector along with an interference rejection combining (IRC) equalizer tailored for FAS. Besides, semi-analytical achievable rates are derived, and are then used to optimize the configuration of FAS. Importantly, we conduct link-level simulations to assess the block error rate (BLER) of the communication links of the OFDM-FAMA system.

Our main contributions are summarized as follows:
\begin{itemize}
\item First, we develop a framework for OFDM-FAMA, facilitating downlink FAMA within the 5G NR architecture. The transmitter adheres to the physical layer procedures of 5G NR \cite{38214}, including channel coding \cite{38212} and OFDM modulation \cite{38211}. At the receiver side, we have developed a port selection mechanism for both training and running stages. Two distinct training strategies are proposed to switch the radio frequency (RF) chains among various fluid antenna ports during the training stage and identify those with the highest signal-to-interference plus noise ratio (SINR). Furthermore, an IRC equalizer is implemented to further mitigate the interference.
\item Three metrics related to the achievable rates of OFDM-FAMA are derived: the average outage rate, the average mutual information (AMI), and the cutoff rate. Moreover, we propose an algorithm to approximate the suboptimal configuration for the FAS at each user, based on the semi-analytical achievable rate. Simulation results reveal that this suboptimal FAS configuration enables OFDM-FAMA to obtain near-optimal system performance.
\item Finally, link-level simulations are conducted considering the tapped delay line (TDL) channel model from the 3rd Generation Partnership Project (3GPP) \cite{38901} to assess the physical layer performance of OFDM-FAMA. Comprehensive BLER results confirm the superior multiplexing gain offered by OFDM-FAMA, particularly in cases with robust channel coding and low spectrum efficiency.
\end{itemize}

{\em Notations:} Scalars are represented by lowercase letters while vectors and matrices are denoted by lowercase and uppercase boldface letters, respectively. Transpose and hermitian operations are denoted by superscript $T$ and $\dag$, respectively. Also, $\lceil \cdot \rceil$ and $\lfloor \cdot \rfloor$ are the ceiling and floor operation, respectively. For a complex scalar $x$, $\lvert x \rvert$ represents its modulus. The notation $*$ denotes the circular convolution operation.

\vspace{-3mm}
\section{Review of FAMA}\label{sec:FAMA}
Before we consider the OFDM-FAMA system, we briefly introduce the FAMA model. In FAMA, an interference channel with $U$ UTs is considered. Each UT is equipped with an $N=(N_1 \times N_2)$-port two-dimensional FAS (2D-FAS) with a physical size of $W_1\lambda \times W_2 \lambda$, where $\lambda$ is the carrier wavelength. Over the 2D space, $N_i$ ports are uniformly distributed along a linear space of length $W_i \lambda$ for $i \in \{1,2\}$. For simplicity, we map the antenna port $(k_1, k_2) \to k: k = k_1\times N_2 +k_2$, where $k_1 \in \{0, \dots, N_1 -1\}$, $k_2 \in \{0, \dots, N_2 -1\}$, and $k \in \{0, \dots, N-1\}$. The received signal at the $k$-th port of user $u$ is modelled as
\begin{equation}
r_k^{(u)}[t] = g_k^{(u,u)} s^{(u)}[t] + \sum_{\substack{\tilde{u}=1\\\tilde{u}\neq u}}^{U} g_k^{(\tilde{u},u)} s^{(\tilde{u})}[t] + \eta_k^{(u)}[t],
\end{equation} 
where $g_k^{(\tilde{u},u)}$ is the fading channel from the $\tilde{u}$-th BS antenna to UT $u$ at the $k$-th port, $\eta_k^{(u)}[t]$ is the zero-mean complex Gaussian noise with variance of $\sigma_\eta^2$, and $s^{(u)}[t]$ is the transmitted symbol for UT $u$ with $\mathbb{E} [|s^{(u)}|^2] = 1$. 

The channels $\{g_k^{(\tilde{u},u)}\}_{\forall k}$ are correlated. With the eigenvalue-based model in \cite{Khammassi2023} and assuming rich scattering, the channel $g_k^{(\tilde{u},u)}$ can be expressed as
\begin{equation}\label{Eq:2Dchan}
g_k^{(\tilde{u},u)} = \sigma^{(\tilde{u},u)} \sum_{l=1}^{N} {\sqrt{\lambda_l} \mu_{k,l}  \alpha_l }\text{,}
\end{equation}
where $\alpha_l \sim \mathcal{CN} (0,1)$, $\lambda_l$ and $\mu_{k,l}$ are obtained from the singular value decomposition (SVD) of the channel covariance matrix $\boldsymbol{\Sigma}$. For simplicity, we set $\sigma^{(\tilde{u},u)} = \sigma, \forall \tilde{u},u$. In this case, the channel covariance matrix can be expressed as
\begin{equation}
\mathbb{E}\left[ \boldsymbol{g}^{(\tilde{u},u)} \left( \boldsymbol{g}^{(\tilde{u},u)} \right) ^ \dag \right] = \sigma^2 \boldsymbol{\Sigma}\text{,}
\end{equation}
where $\boldsymbol{g}^{(\tilde{u},u)}=[ g_1^{(\tilde{u},u)}, \dots, g_N^{(\tilde{u},u)}]^T$, and we have 
\begin{equation}\label{Eq:corr}
\left[\boldsymbol{\Sigma}\right]_{k,l}= J_0 \left( 2\pi \sqrt{\left(\frac{k_1-l_1}{N_1 - 1} W_1\right)^2 + \left(\frac{k_2-l_2}{N_2 - 1} W_2\right)^2}\right),\end{equation}
where $(l_1, l_2) \to l$, and $J_0(\cdot)$ is the zero-order Bessel function of the first order. SVD is carried out as $\boldsymbol{\Sigma} = \boldsymbol{U}\boldsymbol{\Lambda}\boldsymbol{U}^\dag$, where $\boldsymbol{\Lambda} = {\rm diag} ( \lambda_1, \dots, \lambda_N)$, and $[\boldsymbol{U}]_{k,l} = \mu_{k,l}$. 

The key point of FAMA is to select the antenna port(s) with highest SINR at each user for multiple access, i.e., 
\begin{equation}\label{Eq:PortSelect}
k^*_n = \arg \max_{k\backslash \left\{ k^*_0, \dots, k^*_{n-1}\right\}} \Gamma_k,~ n = 0, \dots, N^* -1 \text{,}
\end{equation}
where $N^*$ denotes the number of the selectable antenna ports which depends on the number of RF chains, and $\Gamma_k$ is the average SINR when considering \textit{s}-FAMA \cite{wong2023sFAMA}:
\begin{equation}
\Gamma_k^{\text{\em{s}-FAMA}} = \frac{\lvert g_k^{(u,u)}\rvert ^2}{\sum_{\substack{\tilde{u}=1\\\tilde{u}\neq u}}^{U} \lvert g_k^{(\tilde{u},u)}\rvert ^2 + \sigma_\eta^2}.
\end{equation}

\vspace{-3mm}
\section{Downlink OFDM-FAMA}\label{sec:ofdm-fama}
\begin{figure*}[tbp]
\centering
\includegraphics[width=.8\linewidth]{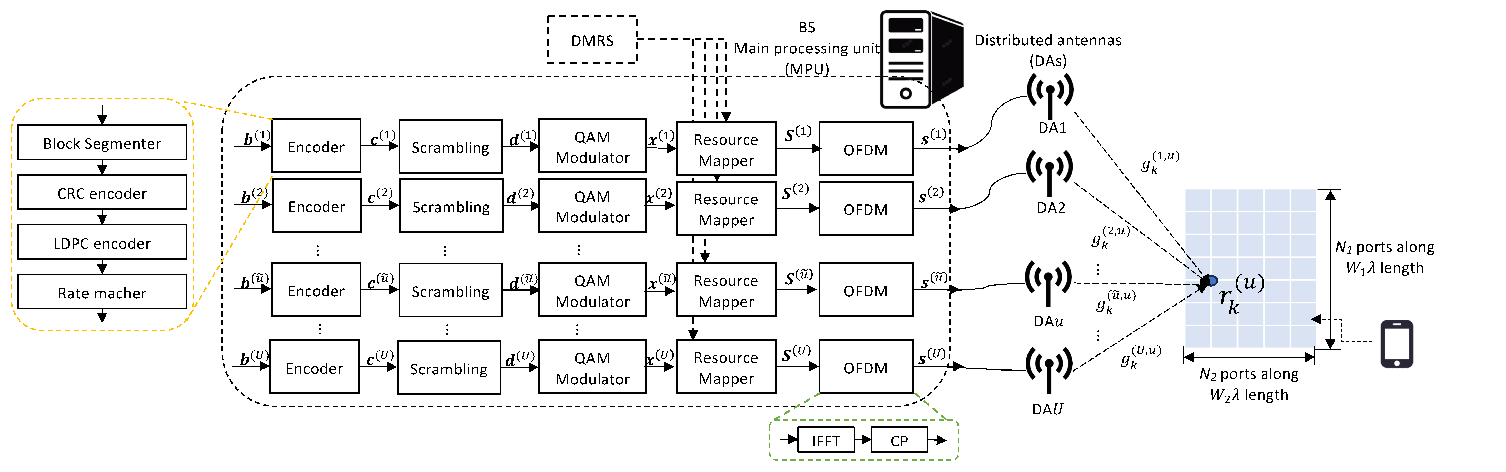}
\caption{System model of the downlink OFDM-FAMA system to a particular UT.}\label{Fig:SysModel}
\vspace{-4mm}
\end{figure*}

Now, consider a downlink OFDM-FAMA system in which the BS communicates to $U$ UTs, each equipped with a FAS. As shown in Fig.~\ref{Fig:SysModel}, the BS has a main processing unit (MPU) and $U$ distributed antennas, each responsible for sending an information-bearing signal to one designated UT. 

At the MPU, the information sequences of $U$ users are processed in parallel with $U$ layers using the same modulation coding scheme (MCS). At the $u$-th layer, the information bit sequence $\boldsymbol{b}^{(u)} = [b_0^{(u)}, \dots, b_{N_a-1}^{(u)} ]$ is encoded to bit sequence $\boldsymbol{c}^{(u)} = [c_0^{(u)},\dots, c_{N_b-1}^{(u)} ]$, where $N_a$ is equal to the transmit block size (TBS), $N_b = N_\text{RE} \times Q_m$ is the number of total transmitted data bits, $N_\text{RE}$ is the number of resource elements for data transmission, and $Q_m$ is the modulation order. A 5G NR encoder is considered, and the encoder includes the block segmenter, the cyclic redundancy check (CRC) encoder, the low-density parity-check (LDPC) encoder, and the rate matcher. The overall code rate can be expressed as 
\begin{equation}
    \text{CR} = N_a/N_b = \text{TBS}/(N_\text{RE}\times Q_m).
\end{equation}
The encoded bit sequence $\boldsymbol{c}^{(u)}$ is scrambled by a user-specific scramble sequence to randomize the date pattern. The scrambled bit sequence $\boldsymbol{d}^{(u)}= [d_0^{(u)},\dots, d_{N_n-1}^{(u)} ]$ is mapped to a symbol sequence $\boldsymbol{x}^{(u)} = [x^{(u)}[0], \dots, x^{(u)}[{N_s-1}]]$, in which $N_s = N_b/Q_m = N_\text{RE}$, where $N_\text{RE}$ is the number of resource elements (RE). This symbol sequence is then mapped to the specific $u$-th antenna and the physical resource blocks (PRBs) for OFDM transmission. The symbol matrix $\boldsymbol{S}^{(u)}$ consists of $N_\text{symb}^\text{subframe} \times (N_\text{PRB}\cdot N_\text{sc}^\text{RB})$ symbols, including data symbols $\boldsymbol{x}^{(u)}$ and the demodulation reference signal (DMRS) symbols, where $N_\text{symb}^\text{subframe}$ is the number of OFDM symbols per subframe, $N_\text{PRB}$ is the number of PRBs, and $N_\text{sc}^\text{RB}$ is the number of subcarriers per resource block (RB). Thus, the overall spectrum efficiency (SE) is calculated as
\begin{equation}\label{eq:SE}
\text{SE} = \text{TBS}/(N_\text{PRB} \times N_\text{sc}^\text{RB} \times N_\text{symb}^\text{Subframe})\text{.}
\end{equation}
The symbol matrix $\boldsymbol{S}^{(u)}$ is then OFDM modulated as $\boldsymbol{s}^{(u)} = [s^{(u)}[0], \dots, s^{(u)}[{N_s^\text{subframe}-1}]]$ and transmitted at the $u$-th distributed antenna, where $N_s^\text{subframe} = N_\text{symb}^\text{subframe} \cdot N_\text{fft} +N_\text{CP}$ is the total number of transmitted symbols per subframe in the time domain, $N_\text{fft}$ is the fast Fourier transform (FFT) size, and $N_\text{CP}$ is the length of the cyclic prefix (CP). 

\begin{figure*}[tbp]
\centering
\includegraphics[width = 0.75\linewidth]{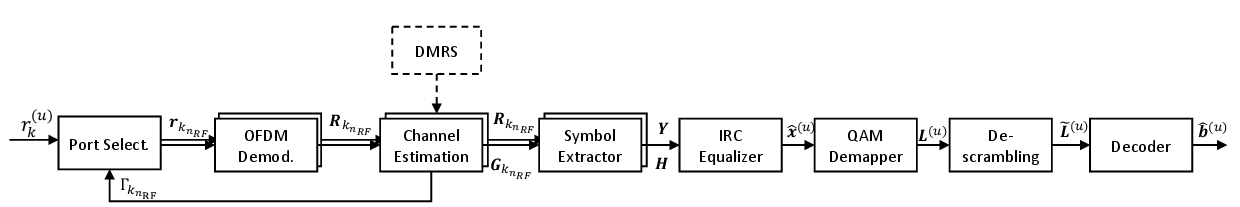}
\caption{Receiving block diagram of the $u$-th UT.}\label{Fig:Receiver}
\vspace{-5mm}
\end{figure*}

The $u$-th UT's receiver with $N_\text{RF}$ RF chains is depicted in Fig.~\ref{Fig:Receiver}, and the received signal at the $k$-th port in the time domain can be expressed as
\begin{equation}
r^{(u)}_k[t] = g_k^{(u, u)}[t] * s^{(u)}[t] +\sum_{{\tilde{u}\neq u}} g_k^{(\tilde u, u)}[t] * s^{(\tilde u)}[t] +\eta_k[t].
\end{equation}
We consider the 3GPP TDL channel model applying the FAS channel correlation matrix, with the power and delay profile specified in \cite{38901}.

With OFDM demodulation, the received signal is transformed, and can be described in the frequency domain as
\begin{equation}\label{Eq:recsymb_freqdom}
    \begin{aligned}
    R_k[n,m] & = G_k^{(u, u)}[n,m] S^{(u)}[n,m] \\
    & +\sum_{{\tilde{u}\neq u}} G_k^{(\tilde u, u)}[n,m] S^{(\tilde u)}[n,m] +Z_k[n,m],
    \end{aligned}
\end{equation}
where $R_k[n,m]$ is the $m$-th FFT bin of the $n$-th OFDM symbol in the received grid $\boldsymbol{R}_k$, $n = 0, \dots, N_\text{symb}^\text{subframe}-1$, $m = 0, \dots, N_\text{PRB}\cdot N_\text{sc}^\text{RB}-1$. Similarly, $G_k^{(\tilde u, u)}[n,m]$, $S^{(\tilde u)}[n,m]$, and $Z_k[n,m]$ are the $(n,m)$-th elements in the channel grid $\boldsymbol{G}_k^{(\tilde u, u)}$, the transmitted symbol grid $\boldsymbol{S}^{(\tilde u)}$, and the Gaussian noise grid $\boldsymbol{Z}_k$, respectively. With FFT operation, these frequency domain signals are given by
\begin{align}
R_k[n,m] &= \text{FFT} (r_k[t],N_\text{fft}),\\
G_k^{(\tilde u, u)}[n,m] &= \text{FFT} (g_k^{(\tilde u, u)}[t],N_\text{fft}),\\
S^{(\tilde u)}[n,m] &= \text{FFT} (s^{(\tilde u)}[t],N_\text{fft}),\\
Z_k[n,m] &= \text{FFT} (\eta_k[t],N_\text{fft}).
\end{align}

\begin{figure}[tbp]
\centering
\vspace{-2mm}
\includegraphics[width = 0.75\linewidth]{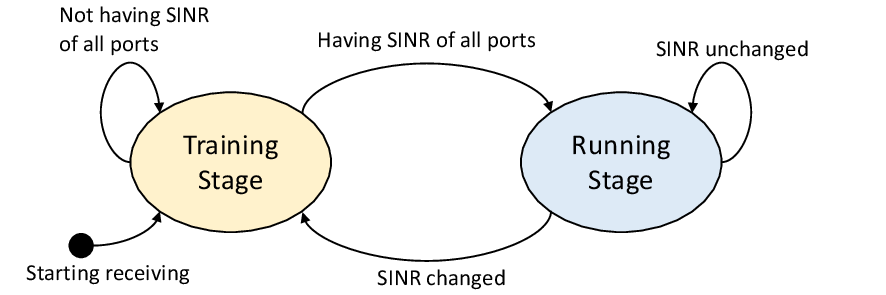}
\caption{Stage chart of the receiver.}\label{Fig:StageChart}
\end{figure}

For FAMA, each UT needs to switch to the best port(s) with the highest SINR(s) \cite{wong2022FAMA}. As shown in Fig.~\ref{Fig:Receiver}, in the OFDM-FAMA system, the average SINR is measured in the frequency domain subframe-by-subframe after channel estimation, while the port selection based on the SINR should be processed in the time domain.\footnote{In the frequency domain, it would be difficult to come up with a metric that can select the proper port affecting multiple subcarriers.} Hence, the measured subframe-average SINRs of antenna ports are fed back to the port selection block, and a mapping between the RF chains and the fluid antenna ports, $(n_\text{RF} \to k): \{k_{n_\text{RF}}\}$, $n_\text{RF} = 0, \dots, N_\text{RF}-1$, is obtained from the measured SINR. But the SINR is incomplete at the beginning of the reception, so we divide the receiving process into training and running stages as shown in Fig.~\ref{Fig:StageChart}. 

During the training stage, the RF chains are switched among the fluid antenna ports subframe by subframe to measure the SINRs of the fluid antenna ports, and the mapping  $\{k_{n_\text{RF}}^{(n_\text{subframe})}\}$ is obtained from the training strategy. In the running stage, the mapping $\{k^*_{n_\text{RF}}\}$ depends upon SINR ordering, in which the antenna ports with the highest SINR are chosen as the active ports. It is worth noting that during the training stage, the data transmission is not interrupted but is transmitted at a relatively lower transmission rate. Meanwhile, the SINR of the selected ports is measured during the running stage, and the system will switch back to the training stage if the SINR changes. The details of the FAS port selection and training strategies will be properly explained in Section \ref{Subsec:TS}.

For the $n_\text{RF}$-th RF chain at a UT receiver, assuming perfect timing synchronization, the received signal at the $k_{n_\text{RF}}$-th port $\boldsymbol{r}_{k_{n_\text{RF}}} = [ r_{k_{n_\text{RF}}}[0], \dots, r_{k_{n_\text{RF}}}[{N_s^\text{subframe}-1}]]$ is OFDM demodulated as the received grid $\boldsymbol{R}_{k_{n_\text{RF}}}$. Using the reference signal, the channel grid can be estimated as $\boldsymbol{G}_{k_{n_\text{RF}}}^{(u,u)}$. The data symbols and their channel are extracted from the received grid and the channel grid as $\boldsymbol{y}_{k_{n_\text{RF}}} = [y_{k_{n_\text{RF}}}[0], \dots, y_{k_{n_\text{RF}}}[{N_s-1}]]$  and $\boldsymbol{h}_{k_{n_\text{RF}}}^{(u,u)} = [{h}_{k_{n_\text{RF}}}^{(u,u)}[0], \dots, {h}_{k_{n_\text{RF}}}^{(u,u)}[{N_s-1}]]$, respectively. From \eqref{Eq:recsymb_freqdom}, the relationship between the extracted data symbol and the channel can be expressed as
\begin{equation}
    y_{k_{n_\text{RF}}}[l] = h_{k_{n_\text{RF}}}^{(u,u)}[l] \cdot x^{(u)}[l] + \eta_{k_{n_\text{RF}}}^{\rm I}[l],
\end{equation}
where the interference and noise signal is given by
\begin{equation}
\eta_{k_{n_\text{RF}}}^{\rm I}[l] = \sum_{\tilde u \neq u} h_{k_{n_\text{RF}}}^{(\tilde u,u)}[l] \cdot x^{(\tilde u)}[l] + \eta[l],
\end{equation}
where $\{h_{k_{n_\text{RF}}}^{(\tilde u,u)}\}$ is the channel coefficient from the $\tilde{u}$-th antenna at the BS to the $k_{n_\text{RF}}$-th port selected by the $n_\text{RF}$-th RF chain of the $u$-th UT which is unknown at the $u$-th UT, and $\eta[l]$ is the complex Gaussian noise with variance of $\sigma_\eta^2$.

Combining all $N_\text{RF}$ RF chains,  the received symbol matrix can be expressed as $\boldsymbol{Y} = [\boldsymbol{y}_{k_0}^T, \dots, \boldsymbol{y}_{k_{N_\text{RF}-1}}^T]^T$, while the channel matrix is $\boldsymbol{H} = [{\boldsymbol{h}_{k_0}^{(u,u)}}^T, \dots, {\boldsymbol{h}_{k_{N_\text{RF}-1}}^{(u,u)}}^T]^T$ . With the IRC equalization, we can have the equalized symbol sequence $\hat{\boldsymbol{x}}^{(u)} = [\hat{x}^{(u)}[0],\dots,\hat{x}^{(u)}[N_s-1]]$. This IRC equalization will be fully introduced in Section \ref{Subsec:IRCEqu}. The soft demapper is then adopted to have the log-likelihood ratio (LLR) sequence $\boldsymbol{L}^{(u)} = [L_0^{(u)}, \dots, L_{N_b-1}^{(u)}]$, and this LLR sequence is descrambled and decoded to restore the information bit sequence.

\vspace{-3mm}
\subsection{Port Selection and Training Strategy}\label{Subsec:TS}
In the port selection process, $N_\text{RF}$ fluid antenna ports with the highest average SINRs are supposed to receive the signals but this can be realized only in the running stage after the SINRs of all ports have been fully acknowledged. When the receiver does not yet know the SINRs of all ports (e.g., at the beginning of the reception), the RF chains should be switched among the fluid antenna ports on a subframe basis to measure the SINRs during reception. This phase is referred to as the training stage. The number of subframes for the training stage, $N_\text{subframe}^\text{TS}$, is determined by the number of ports $N$, the number of RF chains $N_\text{RF}$, and the training strategies employed.

In this paper, we propose two training strategies. \emph{Strategy A} switches all the RF chains, while \emph{Strategy B} switches half of the RF chains and selects the other half from the ports with known SINRs. These two training strategies are described in Algorithms \ref{Alg:TraningStrategyA} and \ref{Alg:TraningStrategyB}, respectively. The training strategies define the port mapping $\{k_{n_\text{RF}}^{(n_\text{subframe})}\}$ for the $n_\text{subframe}$-th subframe during the training stage, and determine the optimal FAS port mapping $\{k^*_{n_\text{RF}}\}$ of the running stage.

\setlength{\textfloatsep}{0pt}
\begin{algorithm} [tbp]
    \caption{Training Strategy A for Port Selection}
    \label{Alg:TraningStrategyA}
    \begin{algorithmic}[1]
        \Require $N$, $N_\text{RF}$;
        \Ensure Training stage $\{k_{n_\text{RF}}^{(n_\text{subframe})}\}$, Running stage $\{k^*_{n_\text{RF}}\}$;
        \State {Interleave the antenna ports as \eqref{Eq:PortInt};}
        \For{$n_\text{subframe} = 0$ to $N_\text{subframe}^\text{TS\emph{A}}-1$}
            \For{$n_\text{RF} = 0$ to $N_\text{RF}-1$}
                \State {$k_{n_\text{RF}}^{(n_\text{subframe})} = \kappa_{(n_\text{subframe}\times N_\text{RF}+n_\text{RF}) \mod N}$;}
            \EndFor
            \State {Switch the RF chains to the $\{k_{n_\text{RF}}^{(n_\text{subframe})}\}$-th fluid antenna ports;}
            \State Measure the SINR of the $\{k_{n_\text{RF}}^{(n_\text{subframe})}\}$-th fluid antenna ports: $\{\Gamma_{k_{n_\text{RF}}^{(n_\text{subframe})}}\}$;
        \EndFor
        \State Sort the SINR list $\boldsymbol{\Gamma} = \{ \Gamma_0, \dots \Gamma_{N-1} \}$;
        \State Select $N_\text{RF}$ ports with largest SINR $\{k^*_{n_\text{RF}}\}$ as \eqref{Eq:PortSelect}.
    \end{algorithmic}
\end{algorithm}
\setlength{\floatsep}{3mm}
\setlength{\textfloatsep}{0pt}
\begin{algorithm} [tbp]
    \caption{Training Strategy B for Port Selection}
    \label{Alg:TraningStrategyB}
    \begin{algorithmic}[1]
        \Require $N$, $N_\text{RF}$;
        \Ensure Training stage $\{k_{n_\text{RF}}^{(n_\text{subframe})}\}$, Running stage $\{k^*_{n_\text{RF}}\}$;
        \State $\Gamma_k = -\inf, \forall k \in \{0,\dots,N-1\}$;
        \State {Interleave the antenna ports as \eqref{Eq:PortInt};}
        \For{$n_\text{subframe} = 0$ to $N_\text{subframe}^\text{TS\emph{B}} -1$}
            \For{$n_\text{RF} = 0$ to $\lfloor N_\text{RF}/2\rfloor -1$}
                \State {$k_{n_\text{RF}}^{(n_\text{subframe})} = \kappa_{ [(n_\text{subframe}+1)\times {\lfloor {N_\text{RF}}/{2}\rfloor}+n_\text{RF}] \mod N}$;}
            \EndFor
            \For{$n_\text{RF} = \lfloor N_\text{RF}/2 \rfloor$ to $N_\text{RF}-1$}
                \If{$n_\text{subframe} = 0$}
                    \State {$k_{n_\text{RF}}^{(n_\text{subframe})} = \kappa_{n_\text{RF}-\lfloor N_\text{RF}/2 \rfloor}$;}
                \Else
                    \State $\begin{aligned}
                        {k_{n_\text{RF}}^{(n_\text{subframe})} = \arg\max_{k \backslash \left\{k_{\lfloor N_\text{RF}/2 \rfloor}^{(n_\text{subframe})}, \dots, k_{n_\text{RF}-1}^{(n_\text{subframe})}\right\}} \Gamma_k}\text{;}
                    \end{aligned}$
                \EndIf
            \EndFor
            \State {Switch to the $\{k_{n_\text{RF}}^{(n_\text{subframe})}\}$-th fluid antenna ports;} 
            \State Measure the SINR: $\{\Gamma_{k_{n_\text{RF}}^{(n_\text{subframe})}}\}$; 
        \EndFor
        \State Sort the SINR list $\boldsymbol{\Gamma} = \{ \Gamma_1, \dots \Gamma_N \}$;
        \State Select $N_\text{RF}$ ports with largest SINR $\{k^*_{n_\text{RF}}\}$ as \eqref{Eq:PortSelect}.
    \end{algorithmic}
    \end{algorithm}

During training, the antenna port will first be interleaved to minimize correlation between the selected antenna ports. The $k$-th port is interleaved to the $\kappa_k$-th port with an interval $\Delta$ as
\begin{equation}\label{Eq:PortInt}
\kappa_k = \begin{cases}
\Delta s + t,    &   \text{if $k < \Delta\lfloor N/\Delta \rfloor$, $k = \Delta t +s$,} \\
                               &   t = 0,1,\dots, \lfloor N/\Delta \rfloor-1,\\
                               &   s = 0,1,\dots, \Delta-1,\\
k,                   &   \text{if $k \geq \Delta\lfloor N/\Delta \rfloor$},
\end{cases}
\end{equation}
where the interval $\Delta = \lceil {N}/{N_\text{RF}} \rceil$ for \emph{Strategy A}, and  $\Delta = \lceil {N}/{\lfloor N_\text{RF}/2 \rfloor} \rceil$ for \emph{Strategy B}. From the $0$-th to the $(N_\text{subframe}^\text{TS}-1)$-th subframe, the RF chains switch among the fluid antenna ports to measure the SINR, $\{\Gamma_{k_{n_\text{RF}}^{(n_\text{subframe})}}\}$. The total number of subframes for training in \emph{Strategy A} is given by
\begin{equation}
    N_\text{subframe}^\text{TS\emph{A}} = \lceil {N}/{N_\text{RF}} \rceil \text{,}
\end{equation}
whereas $N_\text{subframe}^\text{TS}$ for \emph{Strategy B} is given by
\begin{equation}
    N_\text{subframe}^\text{TS\emph{B}} = \lceil {(N-{\lceil {N_\text{RF}}/{2}\rceil})}/{{\lfloor {N_\text{RF}}/{2}\rfloor}} \rceil \text{.}
\end{equation}
Comparing these two numbers, it is evident that \emph{Strategy A} uses fewer subframes during the training stage. After exploring all the fluid antenna ports in the training stage, the optimal port mapping for the running stage can be determined using the FAMA port selection method in \eqref{Eq:PortSelect} by sorting the subframe basis average SINR list $\boldsymbol{\Gamma} = \{\Gamma_0,\dots,\Gamma_{N-1}\}$.

The data transmission performance of the training stage is expected to be lower than that during the running stage. This is because the RF chains may switch to those fluid antenna ports with relatively low SINRs. As a result, the performance of \emph{Strategy B} could surpass that of \emph{Strategy A} after several subframes of training, as \emph{Strategy B} dynamically adjusts the RF chains during the training stage. This will be demonstrated by the link-level simulations in Section \ref{subsec:PerformanceTS}.

\vspace{-3mm}
\subsection{IRC Equalization} \label{Subsec:IRCEqu}
The extracted received symbols among the $N_\text{RF}$ RF chains $\boldsymbol{Y} = [\boldsymbol{y}[0], \dots, \boldsymbol{y}[N_s-1]]$ are combined at the IRC equalizer, where $\boldsymbol{y}[l]$ is the $l$-th column of $\boldsymbol{Y}$. The minimum mean-square-error (MMSE) equalized vector for the $m$-th received symbol $\boldsymbol{y}[l] = [y_{k_0}[l], \dots, y_{k_{N_\text{RF}-1}}[l]]^T$ is given by
\begin{equation}
    \boldsymbol{w}[l] = \boldsymbol{h}^\dag [l] (\boldsymbol{h}[l] \boldsymbol{h} ^\dag [l]+ \hat{\boldsymbol{\Sigma}})^{-1} \text{,}
\end{equation}
where $\boldsymbol{h}[l] = [h_{k_0}^{(u,u)}[l], \dots, h_{k_{N_\text{RF}-1}}^{(u,u)}[l]]^T$ is the channel vector and the $l$-th column vector of the channel matrix $\boldsymbol{H}$, $\hat{\boldsymbol{\Sigma}}$ is the estimated covariance matrix of the interference and noise $\boldsymbol{\eta}^{\rm I}[l] = [\eta^{\rm I}_{k_0}[l], \dots, \eta^{\rm I}_{k_{N_\text{RF}-1}}[l]]^T$. The covariance matrix can be easily estimated as $\hat{\boldsymbol{\Sigma}} = [\left(U-1\right)\sigma^2\boldsymbol{\Sigma} + \sigma^2_\eta \boldsymbol{I}]_{\{k_{n_\text{RF}}\}}$,
or dynamically estimated from the reference signal (RS) as
\begin{equation}
    \begin{aligned}
        \hat{\boldsymbol{\Sigma}} = \frac{1}{N_\text{RS}} \sum \limits_{l=0}^{N_\text{RS}} & (\boldsymbol{y}_\text{RS}[l]-\boldsymbol{h}_\text{RS}[l] x_\text{RS}^{(u)}[l]) \\
        & \times (\boldsymbol{y}_\text{RS}[l]-\boldsymbol{h}_\text{RS}[l] x_\text{RS}^{(u)}[l])^\dag,
    \end{aligned}
\end{equation}
in which $N_\text{RS}$ denotes the number of RS elements, $x_\text{RS}^{(u)}$ is the RS transmitted symbol, $\boldsymbol{y}_\text{RS}$ and $\boldsymbol{h}_\text{RS}$ are the RS received vector and the channel vector extracted from the received grids $\{\boldsymbol{R}_k\}$ and the channel grids $\{\boldsymbol{G}_k^{(u,u)}\}$, respectively. The fixed estimation is simpler, while the dynamical estimation can be more accurate with a sufficiently long RS.

The equalized symbol thus can be expressed as
\begin{equation}
    \begin{aligned}
        \hat{x}^{(u)}[l] & = \beta[l] \boldsymbol{w}[l] \boldsymbol{y}[l] \\
        & = \beta[l] \boldsymbol{w}[l] \boldsymbol{h}[l] x^{u}[l] + \beta[l] \boldsymbol{w}[l] \boldsymbol{\eta}^{\rm I}[l] \text{,}
    \end{aligned}
    \label{Eq:EqualizedSym}
\end{equation}
where $\beta[l] = 1/\lvert \boldsymbol{w}[l]\boldsymbol{h}[l]\rvert$ is the normalized factor. 

In the equalized symbol \eqref{Eq:EqualizedSym}, $\beta\boldsymbol{w}\boldsymbol{h}x^{(u)}$ is the desired signal part, while $\beta\boldsymbol{w} \boldsymbol{\eta}^{\rm I}$ is the noise and interference part. Thus, the average SINR $\overline{\Gamma}$ of the subframe is given by
\begin{equation} \label{eq:averageSINR}
    \begin{aligned}
        \overline{\Gamma} & = \frac{\mathbb{E}_l\left\{{\lvert \beta[l] \boldsymbol{w} [l]\boldsymbol{h} [l] x^{(u)}[l] \rvert ^2}\right\}}{\mathbb{E}_l\left\{ \lvert\beta[l] \boldsymbol{w} [l] \boldsymbol{\eta}^{\rm I} [l]\rvert^2\right\}} \\
        & = {1}/\mathbb{E}_l\left\{ {\lvert\beta[l] \boldsymbol{w} [l] \boldsymbol{\eta}^{\rm I} [l]\rvert^2}\right\}.
    \end{aligned}
\end{equation}
This average SINR is the metric for performance evaluation.

\vspace{-3mm}
\subsection{Achievable Rate and FAS Configuration}
The outage probability is defined as the occurrence of the average SINR $\overline{\Gamma}$ being less than the target SINR $\Gamma$. In other words, the outage probability for the $u$-th UT is given by 
\begin{equation}
    p_\text{out} \triangleq \text{Prob}\left(\overline{\Gamma} = \mathbb{E}\left\{ {1}/{\lvert\beta[l] \boldsymbol{w} [l] \boldsymbol{\eta}^{\rm I} [l]\rvert^2}\right\} < \Gamma \right).
\end{equation}
For a specific MCS with $\text{SE}$ calculated in \eqref{eq:SE}, the target SINR for outage probability evaluation can be estimated as
\begin{equation} \label{eq:targetSINR}
    \Gamma = 2^{\text{SE}}-1.
\end{equation}
The average outage rate can be evaluated by
\begin{equation} \label{Eq:OutageRate}
C_{\Gamma} = U(1-\text{Prob} (\overline{\Gamma}<\Gamma)) \times \frac{\text{TBS}}{N_\text{PRB} N_\text{sc}^\text{RB} N_\text{symb}^\text{Subframe}}.
\end{equation}
This corresponds to the case in which the BS transmits a fixed MCS with TBS information bits to the users. The multiplexing gain, $M$, is the capacity scaling factor given by
\begin{equation} \label{Eq:MultiplexingGain}
    M = \frac{C_{\Gamma}}{\log_2(1+\Gamma)}=U(1-\text{Prob} (\overline{\Gamma}<\Gamma)).
\end{equation}

In addition to the outage rate in \eqref{Eq:OutageRate}, the system rate can also be assessed through the AMI and the cutoff rate. The AMI per UT of OFDM-FAMA can be calculated as
\begin{equation} \label{Eq:AMI}
    C_B = Q_m + \sum_{i=1}^{Q_m} \mathbb{E}_{b,\boldsymbol{y},\boldsymbol{h}} \left[{\log_2 {\frac{\sum_{x\in \boldsymbol{\chi}_i^{(b)}} p_{\boldsymbol{h}}(\boldsymbol{y}|x)}{\sum_{x\in \boldsymbol{\chi}} p_{\boldsymbol{h}}(\boldsymbol{y}|x)} }}\right],
\end{equation}
where the PDF $p(\boldsymbol{y}|x)$ is given by IRC as
\begin{equation}
p_{\boldsymbol{h}}(\boldsymbol{y}|x)  = \frac{1}{\pi^{N_\text{RF}}\lvert \hat{\boldsymbol{\Sigma}} \vert } 
\exp\left[-(\boldsymbol{y}-\boldsymbol{h}x)^\dag \hat{\boldsymbol{\Sigma}} ^{-1}(\boldsymbol{y}-\boldsymbol{h}x)\right].
\end{equation}
Also, the cutoff rate can be obtained from the Bhattacharyya bound on the average bit-error probability in the absence of coding. The average Bhattacharyya factor is given by
\begin{equation}
    B = \frac{1}{Q_m} \sum_{i=1}^{Q_m} \mathbb{E}_{b,\boldsymbol{y},\boldsymbol{h}} \left[ \sqrt{\frac{\sum_{z \in \boldsymbol(\chi)_i^{(\overline{b})}} p_{\boldsymbol{h}}(\boldsymbol{y}|x) }{\sum_{z \in \boldsymbol(\chi)_i^{(b)}} p_{\boldsymbol{h}}(\boldsymbol{y}|x) }}\right].
\end{equation}
With that, the cutoff rate $C_R$ can be written as
\begin{equation} \label{Eq:CutoffRate}
    C_R = Q_m(1-\log_2(B+1)).
\end{equation}

Now let us discuss the configuration of the FAS equipped at the UTs. Research in \cite{Khammassi2023,New2023fluid} indicates that for a fixed FAS physical size $W = [W_1,W_2]$, the achievable rate may remain similar beyond a certain threshold $N^*$. Further, \cite{New2023fluid} proposed a method to approximate $N^*$ for a given size $W$ by analyzing the eigenvalues $\{\lambda_n\}$ of the channel covariance matrix $\boldsymbol{\Sigma}$. While this method is proven useful for FAS transmission with 1D-FAS, how the result can be extended to our OFDM-FAMA network utilizing the 2D-FAS at the receiver is unknown.

To address this, we present Algorithm \ref{Alg:approxN} that approximates the suboptimal configuration $N^*$ by directly evaluating the rate. For simplicity, a symmetrical 2D-FAS is considered, i.e., $W_1 = W_2$ and $N_1 = N_2$. The achievable rate $C^{(N_1 \times N_2)}$ in Algorithm \ref{Alg:approxN} can refer to the outage rate in \eqref{Eq:OutageRate}, the AMI in \eqref{Eq:AMI}, or the cutoff rate in \eqref{Eq:CutoffRate}. This algorithm determines the suboptimal $N^*$ where OFDM-FAMA provides only a marginal improvement in the rate when $N>N^*$, given a threshold $\epsilon_C$. Other system configurations, including the normalized physical size $W = [W_1, W_2]$, the number of supported UTs $U$, and the number of RF chains $N_\text{RF}$, need to be specified in the algorithm to calculate the achievable rate. If the AMI or cutoff rate is chosen as the metric, the modulation constellation $\boldsymbol{\chi}$ should be specified. If the outage rate is chosen as the criterion, the suboptimal $N^*$ will depend on the target SINR $\Gamma$. 

\begin{algorithm}[tbp]
\caption{Approximating FAS configuration $N^*$}\label{Alg:approxN}
	\begin{algorithmic}
	\Require {$W$, $N_\text{RF}$, $U$, threshold $\epsilon_C$, ($\boldsymbol{\chi}$ or $\Gamma$)}
	\Ensure {$N^*$}
	\State $N_1 = N_2 = \lfloor \sqrt{N_\text{RF}} \rfloor$;
	$C^{\lfloor \sqrt{N_\text{RF}} \rfloor \times \lfloor \sqrt{N_\text{RF}} \rfloor} = 0$;
	\While{$C^{(N_1 \times N_2)} -  C^{(N_1-1) \times (N_2-1)} > \epsilon_C$}
		\State $N_1 = N_1 + 1$;
		\State $N_2 = N_2 + 1$;
		\State Calculate $C^{(N_1\times N_2)}$ for $N = N_1 \times N_2$;
    \EndWhile
    \State $N^* = N_1 \times N_2$;
    \end{algorithmic}
\end{algorithm}
\setlength{\floatsep}{-2pt}

The complexity of the proposed method is greater than that of the approach in \cite{New2023fluid}, since it employs Monte-Carlo numerical integration to calculate the achievable rates. However, this algorithm is a one-time task. Therefore, its complexity does not adversely affect the overall process of OFDM-FAMA transmission, making this high complexity acceptable.

\vspace{-2mm}
\section{Semi-Analytical Performance Evaluation}\label{sec:theoAna}
This section presents semi-analytical results to predict the performance of OFDM-FAMA during the running stage. Given the mathematical difficulties, Monte-Carlo simulations with $10,000$ independent channel realizations are used to compute these metrics. The analysis is conducted under the assumption of a rich scattering block fading channel with perfect channel estimation. As the evaluations pertain specifically to the running stage, we assume that the FAS at the UT is capable of identifying the ports with the highest average SINRs. Table \ref{Tab:SimPara} outlines the parameters and corresponding values considered in the evaluations. The carrier frequency is set as $5$ GHz, resulting in a wavelength of $\lambda = 6$ cm. The normalized antenna sizes are $W$ of $[2,2]$ and $[5,5]$, leading to the actual physical dimensions of $12~\text{cm}\times 12~\text{cm}$, and $30~\text{cm}\times 30~\text{cm}$, respectively. 

\begin{table}[t]
    \begin{center}
        \vspace{-2mm}
        \caption{Simulation Parameters}
        \label{Tab:SimPara}
        \begin{tabular}{l|l}
            \hline
            \textbf{Parameter}                  & \textbf{Value} \\ \hline\hline
            Normalized size of FAS $[W_1,W_2]$  & $[2,2]$, $[5,5]$  \\ \hline
            Number of ports $N_1 \times N_2$    & From $2\times 2$ to $15\times 15$      \\ \hline
            Number of RF chains $N_\text{RF}$   & $2$, $4$, $16$    \\ \hline
            Bandwidth                           & $1.4$ MHz         \\ \hline
            Number of PRBs $N_\text{PRB}$       & $6$               \\ \hline
            Number of subcarriers per RB $N_\text{sc}^\text{RB}$    & $12$  \\ \hline
            Subcarrier spacing $\Delta f$       & $15$ kHz          \\ \hline
            Symbol duration                     & $66.7$ {\textmu}s \\ \hline
            CP duration                         & $4.7$ {\textmu}s  \\ \hline
            FFT Size $N_\text{fft}$             & $128$             \\ \hline
            \makecell[l]{Number of symbols per subframe \\
             $N_\text{symb}^\text{subframe}$}   & $14$              \\ \hline
            Number of resource elements $N_\text{RE}$   & $936$     \\ \hline
            Signal-to-noise ratio (SNR) $\sigma^2/\sigma_\eta^2$ & $35$ dB \\ \hline
            Modulation coding scheme            & MCSs in \cite[Table 5.1.3.1-1]{38214}    \\ \hline
            Channel model           & \makecell[l]{Block fading channel,    \\
                                        TDL-C channel \cite{38901}} \\ \hline
            Delay spread                        & $30$ ns           \\ \hline
            Maximum Doppler frequency           & $0$, $30$, $300$ Hz\\ \hline
        \end{tabular}
    \end{center}
\end{table}

\vspace{-3mm}
\subsection{Achievable Rate}

\begin{figure}[t]
\vspace{-2mm}
\centering
\includegraphics[width = 0.8\linewidth]{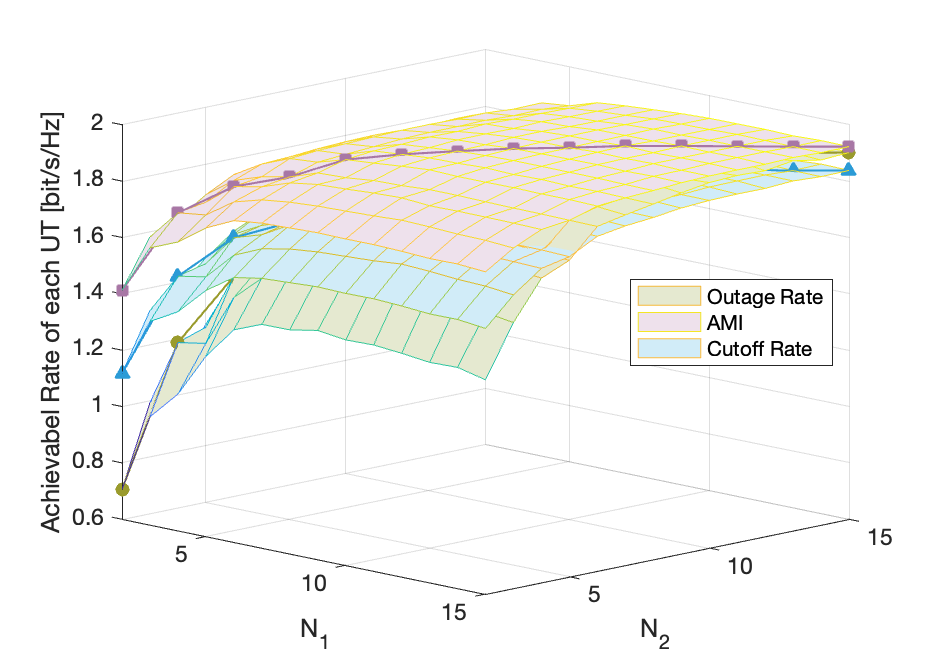}
\caption{Achievable rates of each UT with target SINR $\Gamma =5$ dB, when $W = [2,2]$, $N_\text{RF}=4$, and $U=6$.}\label{Fig:Ratevs2DN}
\end{figure}

Fig.~\ref{Fig:Ratevs2DN} presents the results of the outage rate $C_\Gamma$, the mutual information $C_B$, and the cutoff rate $C_R$, against the 2D number of ports, $N_1$ and $N_2$, in the FAS-equipped UT in OFDM-FAMA, with $U=6$ users. The physical size of FAS equipped at the UT is $2\lambda \times 2\lambda$. The modulation constellation utilized for evaluation of AMI and cutoff rate is quadrature phase shift keying (QPSK). To ensure a fair comparison, an uncoded QPSK system with target SINR of $\Gamma = 5$ dB is considered in the outage evaluation. The results indicate that the performance is symmetric. Consequently, we will maintain the symmetry condition (i.e., $N_1=N_2$ and $W_1=W_2$) throughout the rest of the simulations. In addition, an increase in the number of ports correlates with an enhancement in achievable rates, with a notable improvement observed when $N$ is small, while the enhancement becomes marginal with large $N$. 

\begin{figure*}[t]
    \centering
    \subfloat[$\Gamma = 5$ dB, $N_\text{RF}=4$]{\includegraphics[width=0.32\linewidth]{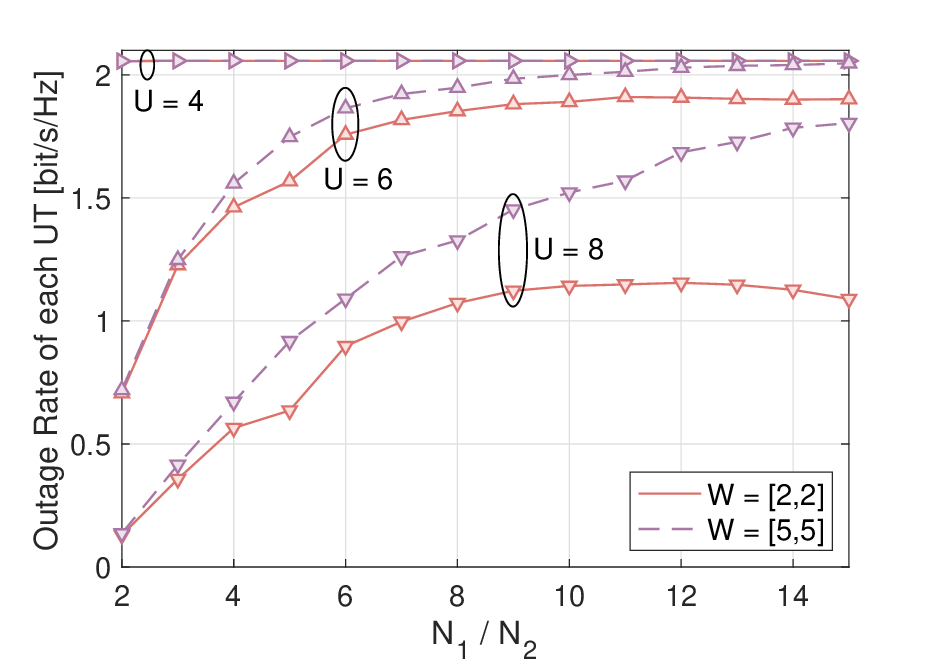}\label{SubFig:ORvsN_NRF4}}
    \subfloat[QPSK, $N_\text{RF}=4$]{\includegraphics[width=0.32\linewidth]{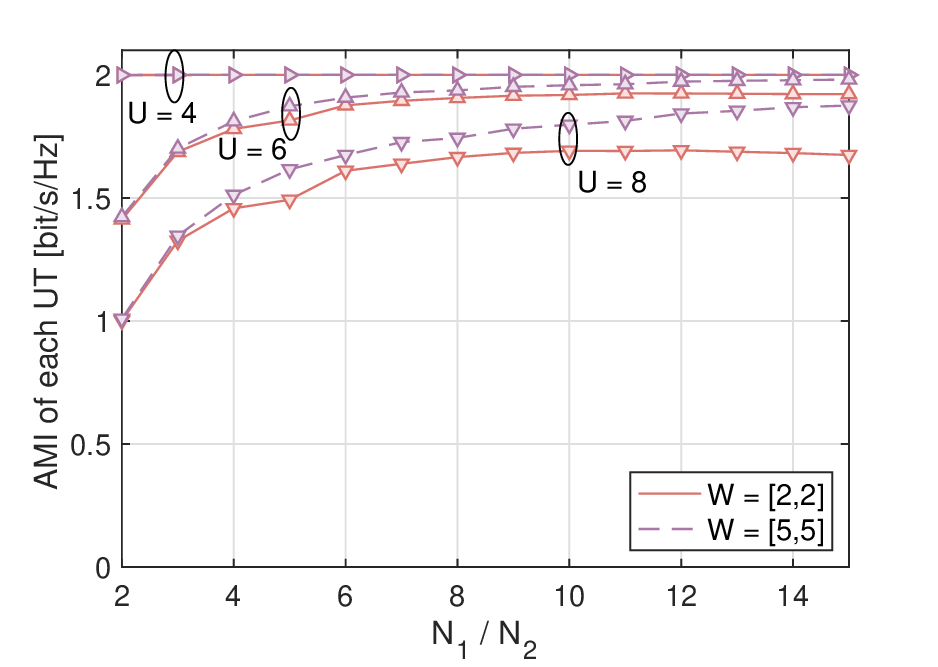}\label{SubFig:AMIvsN_NRF4}}
    \subfloat[QPSK, $N_\text{RF}=4$]{\includegraphics[width=0.32\linewidth]{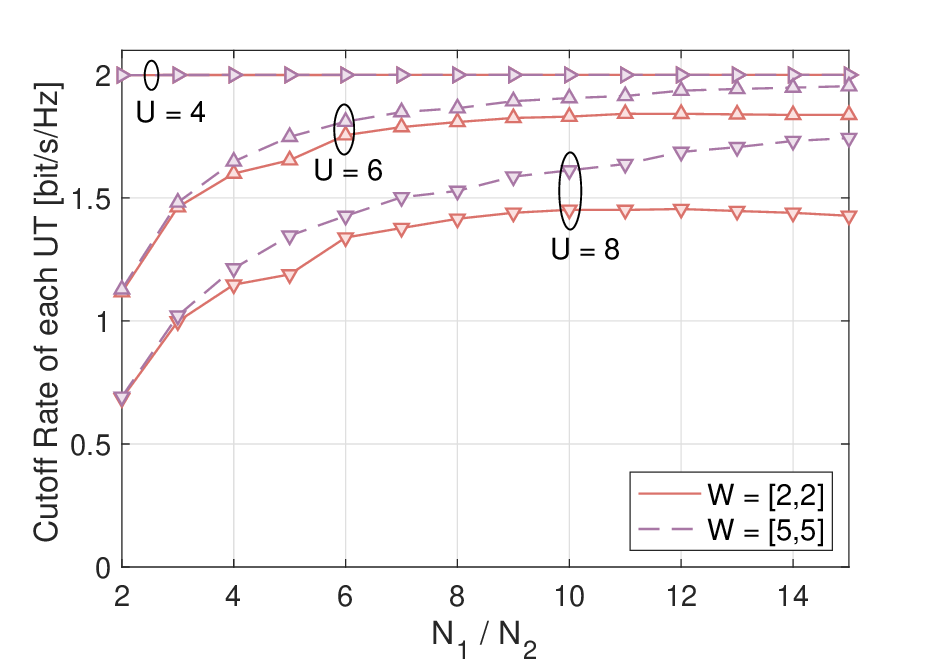}\label{SubFig:CRvsN_NRF4}}\\\vspace{-3mm}
    \subfloat[$\Gamma = 5$ dB, $N_\text{RF}=16$]{\includegraphics[width=0.32\linewidth]{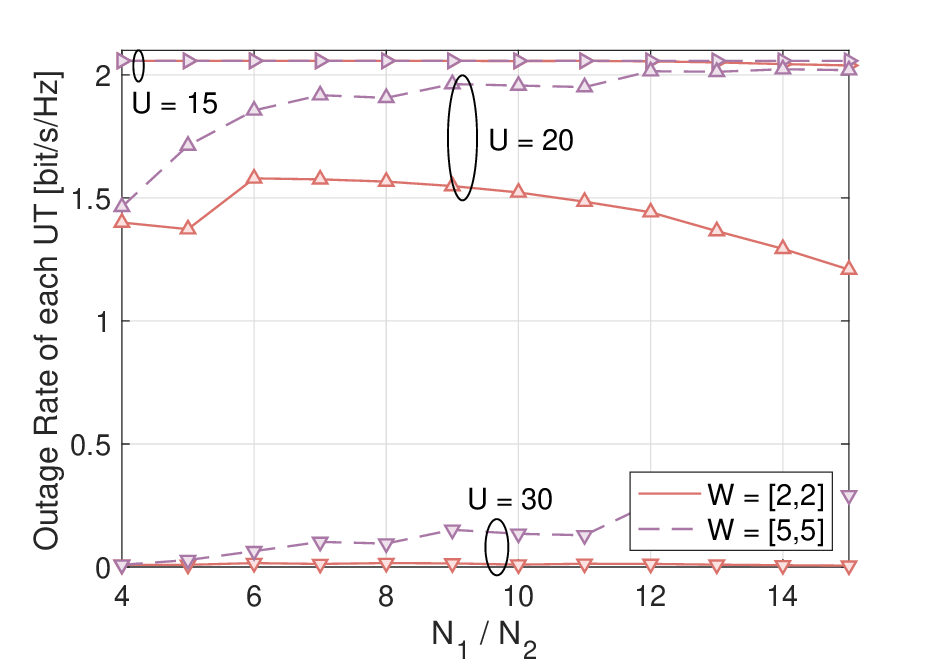}\label{SubFig:ORvsN_NRF16}}
    \subfloat[QPSK, $N_\text{RF}=16$]{\includegraphics[width=0.32\linewidth]{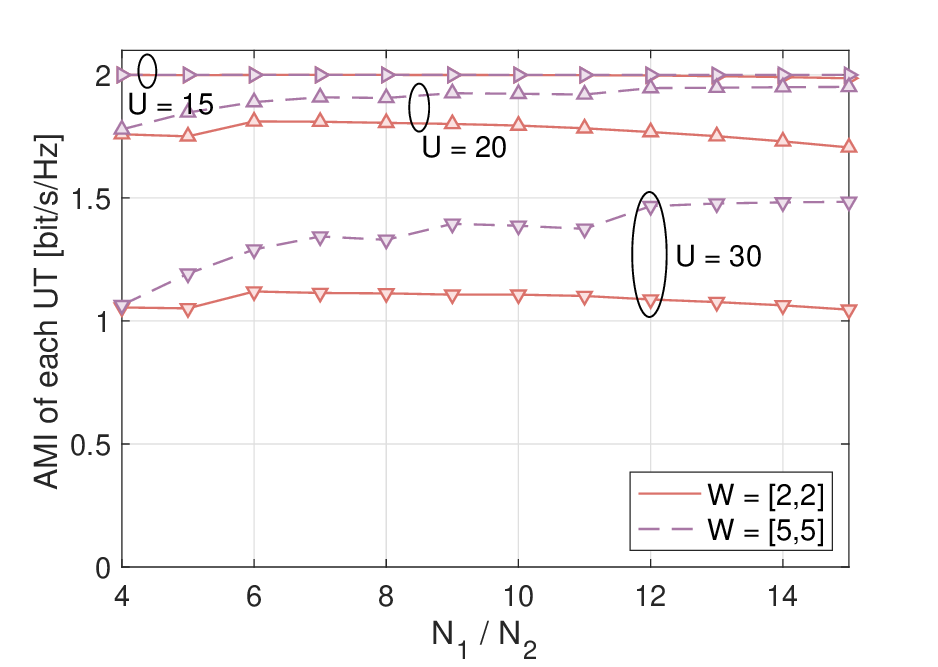}\label{SubFig:AMIvsN_NRF16}}
    \subfloat[QPSK, $N_\text{RF}=16$]{\includegraphics[width=0.32\linewidth]{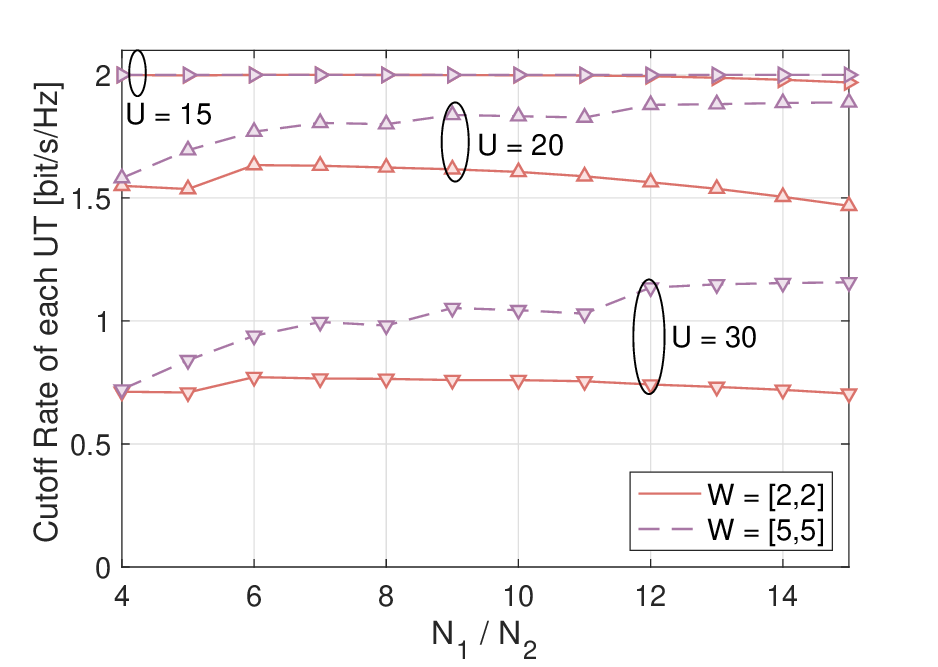}\label{SubFig:CRvsN_NRF16}}
    \caption{Achievable rates of each UT against $N_1$ (or $N_2$), with different $N_\text{RF}$, $W$, and $U$, and the target SINR $\Gamma = 5$ dB.}\label{Fig:Ratevs1DN}
\vspace{-4mm}
\end{figure*}

With $N_1 = N_2$, different rate results against a range of parameters are shown in Fig.~\ref{Fig:Ratevs1DN}. The modulation constellation and the target SINR are the same as that in Fig.~\ref{Fig:Ratevs2DN}. Figs.~\ref{Fig:Ratevs1DN}\ref{sub@SubFig:ORvsN_NRF4}--\ref{Fig:Ratevs1DN}\ref{sub@SubFig:CRvsN_NRF4} consider the case when $N_\text{RF}=4$, whereas Figs.~\ref{Fig:Ratevs1DN}\ref{sub@SubFig:ORvsN_NRF16}--\ref{Fig:Ratevs1DN}\ref{sub@SubFig:CRvsN_NRF16} pertain to $N_\text{RF}=16$. When $N_\text{RF} = 4$ and $U=4$, the rate is limited by the modulation order $Q_m = 2$ bit/s/Hz, or the Shannon capacity of target SINR $\Gamma$, calculated as $\log_2(1+\Gamma) = 2.06$ bit/s/Hz. In this case, error-free transmission is achievable, even when $N = 2\times 2$, which degrades the antenna configuration at each UT to a fixed-position antenna (FPA). In addition, we can observe that the achievable rates become saturated when $N_\text{RF}=16$ and $U=15$, allowing for error-free transmission even under FPA condition ($N = 4\times 4$). As the number of UTs $U$ increases, the achievable rates have a positive correlation with the number of ports, $N=N_1\times N_2$. The increasing of the rates is significant when $N$ is small, but becomes marginal when $N$ grows large. The saturated rate for the system accommodating a large number of users ($U=6$ or $8$ for $N_\text{RF}=4$ and $U=20$ or $30$ for $N_\text{RF}=16$) is lower than $2$ bit/s/Hz. Furthermore, when $N_\text{RF} = 16$ and $W = [2,2]$, the increasing of the achievable rate is marginal at the beginning, since the FPA with $N = 4\times 4$ almost reaches the correlation saturation point over this physical size.

Note that when $N_\text{RF} = 16$ and $U = 30$, the outage rate nears $0$ bit/s/Hz. The reason is that uncoded QPSK struggles to transmit signals for such a high number of users, underscoring the necessity for robust channel coding when designing a system to support massive UTs. An intriguing observation is that when the number of antenna ports is exceedingly large, the achievable rate will slightly decrease with the increase of $N_1$ (or $N_2$), because of the correlation among the ports. This suggests that an excessive increase in the number of ports might adversely affect the overall performance. 

Rate comparison with different physical sizes $W$ indicates that a larger size of FAS at the UTs contributes to a higher achievable rate, with the disparity widening as the number of ports increases, because the saturated $N^*$ is smaller for a small size $W$. On the other hand, the achievable rates of the system with FPA remain consistent when $W = [2,2]$ and $[5,5]$, which means the physical size does not significantly influence the performance of the system employing FPA, while it does affect the performance of OFDM-FAMA utilizing FAS.

\vspace{-3mm}
\subsection{Suboptimal FAS Configuration}

\begin{table}[t]
\begin{center}
\caption{The suboptimal $N^*$ using Algorithm \ref{Alg:approxN} for $W$ and $N_\text{RF}$ with different $C^{(N_1\times N_2)}$, where $\epsilon_C = 0.02$ bit/s/Hz} \label{Tab:optimalN}\vspace{-2mm}
\begin{tabular}{c|c|c|c|c|c|c}
\hline
\multirow{2}{*}{$W$} & \multirow{2}{*}{$N_\text{RF}$} & \multirow{2}{*}{$U$} & \multicolumn{3}{c|}{criterion $C^{(N_1\times N_2)}$} &  Algorithm 1\\ \cline{4-6}
& & & $C_\Gamma$ & $C_B$   & $C_R$ & in \cite{New2023fluid}\\ 
\hline\hline
\multirow{4}{*}{$[2,2]$} & \multirow{2}{*}{$4$} & 6 & $9\times 9$    & $7 \times  7$    & $7 \times 7$ & \multirow{4}{*}{$5 \times 5$}\\ 
\cline{3-6}
& & 8 & $9 \times  9$   & $8 \times 8$ & $8 \times 8$    & \\
\cline{2-6}
& \multirow{2}{*}{$16$} & 20    & $6 \times 6$ & $6 \times 6$ & $6 \times 6$ & \\
\cline{3-6}
& & 30 & -- & $6 \times 6$ & $6 \times 6$ & \\ 
\hline
\multirow{4}{*}{$[5,5]$} & \multirow{2}{*}{$4$} & 6 & $9 \times  9$     & $7 \times 7$ & $7 \times 7$ & \multirow{4}{*}{$8 \times 8$}\\ 
\cline{3-6}
& & 8 & $14 \times 14$ & $12 \times 12$ & $12 \times 12$ & \\ 
\cline{2-6}
& \multirow{2}{*}{$16$} & 20    & $12 \times 12$ & $6 \times 6$ & $12 \times 12$ & \\ 
\cline{3-6}
& & 30 & $12 \times 12$ & $12 \times 12$ & $12 \times 12$ & \\ 
\hline
\end{tabular}
\end{center}
\end{table}

Based on the rates in Fig.~\ref{Fig:Ratevs1DN}, we employ Algorithm \ref{Alg:approxN} to find the suboptimal $N^*$ for FAS at each UT with different $W$ and $N_\text{RF}$. The threshold for the rate $\epsilon_C$ is set as $0.02$ bit/s/Hz.  With the exception of certain abrupt saturation instances, the suboptimal $N^*$ for various configurations are demonstrated in Table \ref{Tab:optimalN}. The calculation is not performed when $C^{(N_1\times N_2)} = C_\Gamma$, $W = [2,2]$, $N_\text{RF} = 16$, and $U=30$, as the outage rate remains nearly zero in this case. Furthermore, the approximation results of \cite[Algorithm 1]{New2023fluid} are included in the table for comparison. Given that \cite[Algorithm 1]{New2023fluid} is specifically designed for 1D-FAS, we regard the $\lceil \sqrt{N^*_\lambda} \times \sqrt{N^*_\lambda}\rceil$ as the suboptimal solution for 2D-FAS at each UT, where $N^*_\lambda$ is the suboptimal number of ports obtained from \cite[Algorithm 1]{New2023fluid}. As shown in Table \ref{Tab:optimalN} and Fig.~\ref{Fig:Ratevs1DN}, the achievable rates at each UT of the OFDM-FAMA system demonstrate considerable improvements when $N$ exceeds the suboptimal solution of \cite[Algorithm 1]{New2023fluid}. Thus, it is wise to consider the pragmatic achievable rates during the design phase of the FAS configuration in the OFDM-FAMA system, despite the additional complexity.

\vspace{-3mm}
\subsection{Multiplexing Gain}\label{subsec:MG}
\begin{figure}
\centering
\subfloat[$W = 2\times 2$]{\includegraphics[width=0.9\linewidth]{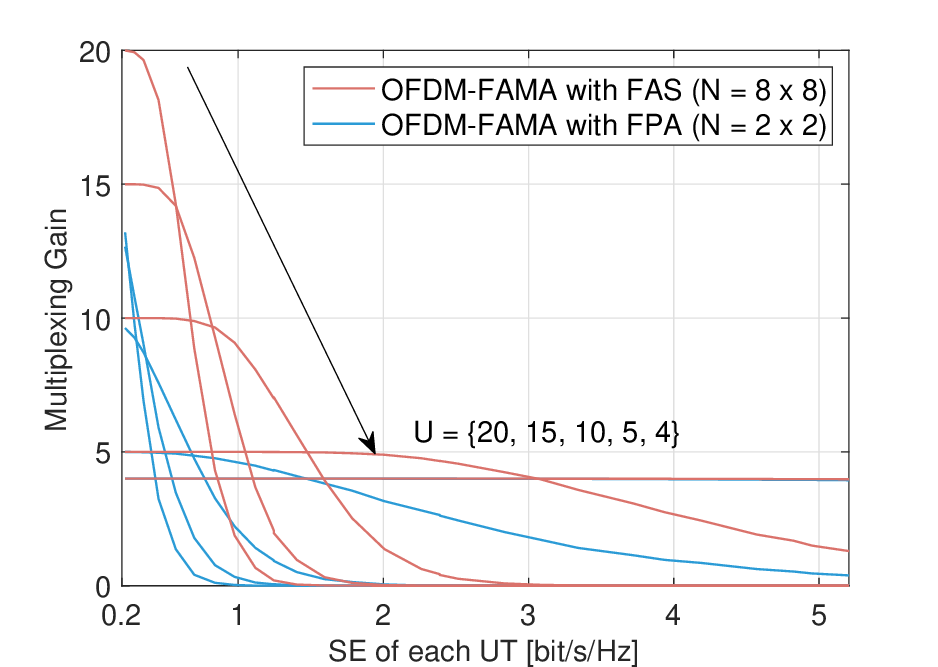}\label{SubFig:MGvsSE_W2x2}}\\
\vspace{-3mm}
\subfloat[$W = 5\times 5$]{\includegraphics[width=0.9\linewidth]{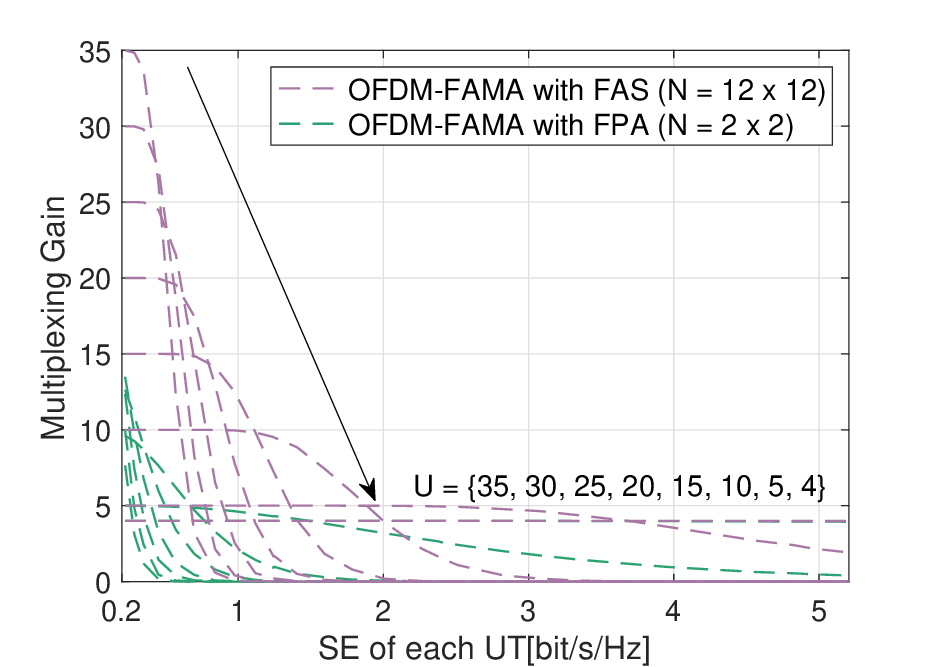}\label{SubFig:MGvsSE_W5x5}}
\caption{Multiplexing gains of OFDM-FAMA against the SE, with the number of RF chains $N_\text{RF} = 4$ and the normalized antenna size: (a) $W = 2\times 2$, and (b) $W = 5\times 5$.}\label{Fig:MGvsSE}
\end{figure}

The multiplexing gains of OFDM-FAMA are illustrated in Fig.~\ref{Fig:MGvsSE}, relative to the SE of each UT, with the number of RF chains set as $N_\text{RF} = 4$. We assume that an adequate number of antenna ports are allocated for the FAS at each user, with the number of antenna ports determined by Algorithm \ref{Alg:approxN}, utilizing $U = 2 N_\text{RF}$ and $C^{(N_1\times N_2)} = C_R$. Consequently, there are $N = 8\times 8$ antenna ports occupying a physical size of $2\lambda \times 2 \lambda$, and $N = 12\times 12$ ports within $5\lambda \times 5\lambda$ physical size for the FAS. The evaluation of multiplexing gain is conducted as per \eqref{Eq:MultiplexingGain} with the target SINR defined by \eqref{eq:targetSINR}. 

An increase in SE implies an increase in the target SINR $\Gamma$, which subsequently leads to the increase of outage probability, causing a rapid decrease in multiplexing gain beyond the transmission capability. In addition, we can observe that the multiplexing gain decreases mildly with a low $U$, indicating that the transmission capability of each UT increases as $U$ decreases. The reason is that the average SINR $\overline{\Gamma}= \lvert\boldsymbol{w}\boldsymbol{y}\rvert ^2 / \lvert\boldsymbol{w}\boldsymbol{\eta}^{\rm I}\rvert ^2 $ over the block fading channel is inversely related to the number of interferers, ($U-1$). At a low SE, the multiplexing gain in \eqref{Eq:MultiplexingGain} is limited by $U$, as the target SINR is low and $p_\text{out} \to 0$. This observation implies that the system with robust coding can support a greater number of UTs simultaneously.

Additionally, the multiplexing gains of the system with FPA at each UT are also presented in Fig.~\ref{Fig:MGvsSE}, with $N = 2\times 2$ for $N_\text{RF} = 4$. It can be observed that the OFDM-FAMA system exhibits superior performance compared to the system with FPA, particularly at low SE and large $U$. But the multiplexing gains converge and become similar as the SE increases and $U$ decreases. Notably, when SE is lower than $0.5$ bit/s/Hz, the multiplexing gain of the OFDM-FAMA system approaches nearly double that of the system with FPA. This illustrates that the OFDM-FAMA system is particularly suitable for scenarios characterized by a large number of UTs needing to transmit small information packets within the limited spectrum.

\vspace{-2mm}
\section{Link-Level Simulation Results}\label{sec:LLS}
Link-level simulations are conducted to evaluate the physical layer performance of the OFDM-FAMA system, where $U$ streams are processed in parallel and transmitted by $U$ antennas simultaneously over block fading and TDL-C channel environments. At the $u$-th UT, a 2D-FAS with $N = N_1\times N_2$ ports over the normalized size of $W = [W_1,W_2]$ is considered. For the running stage, it is assumed that the FAS can identify and receive signals from the desirable antenna ports. Perfect time synchronization and channel estimation are assumed. The remaining assumptions and parameters we used in the link-level simulations are summarized in Table \ref{Tab:SimPara}. Considering the DMRS overhead in a PRB as $N_\text{DMRS} = 12$, the total number of REs allocated for physical downlink share channel (PDSCH) in a subframe is $N_\text{RE} = 936$.

We first present the stationary performances of the OFDM-FAMA system during the running stage over the rich-scattering block fading and the TDL-C channels. Then we evaluate the OFDM-FAMA system under mobility situations by link-level simulations over TDL-C channels in low-speed scenarios with $f_D = 30$ Hz and high-speed scenarios with $f_D = 300$ Hz. Finally, the performance of training stage is evaluated.

\vspace{-2mm}
\subsection{BLER Performance}

\begin{figure}[t]
\centering
\includegraphics[width = 0.85\linewidth]{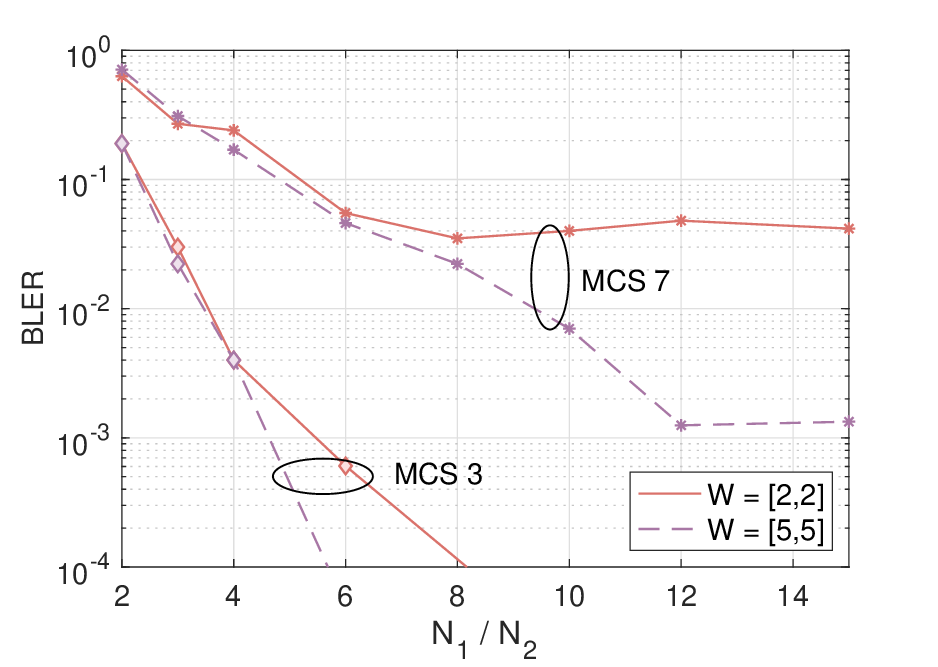}
\caption{BLER against the number of antenna ports $N_1$ (or $N_2$), with $N_\text{RF} = 4$ and $U = 8$, over block fading channels.}\label{Fig:NvsBLER}
\end{figure}

Fig.~\ref{Fig:NvsBLER} demonstrates the BLER results against the number of antenna ports of FAS at each user, utilizing MCS 3 and 7, with $N_\text{RF}= 4$ and $U = 8$. MCS 3 and MCS 7 facilitate transmission rate of $0.5$ and $1.0$ bit/s/Hz for each user, respectively. Block fading channel with rich scattering Rayleigh fading is considered in the results of this figure. As expected, BLER decreases with an increase in the number of antenna ports, $N$. However, an error floor exists when $N$ is large because of the correlation among the antenna ports within a limited space. This error floor occurs at the lower BLER when the physical size of FAS for each user is increased, or when the channel coding is more robust. These findings are consistent with the semi-analytical evaluations of the rates in Fig.~\ref{Fig:Ratevs1DN}.

\begin{figure}[t]
\centering
\includegraphics[width = 0.9\linewidth]{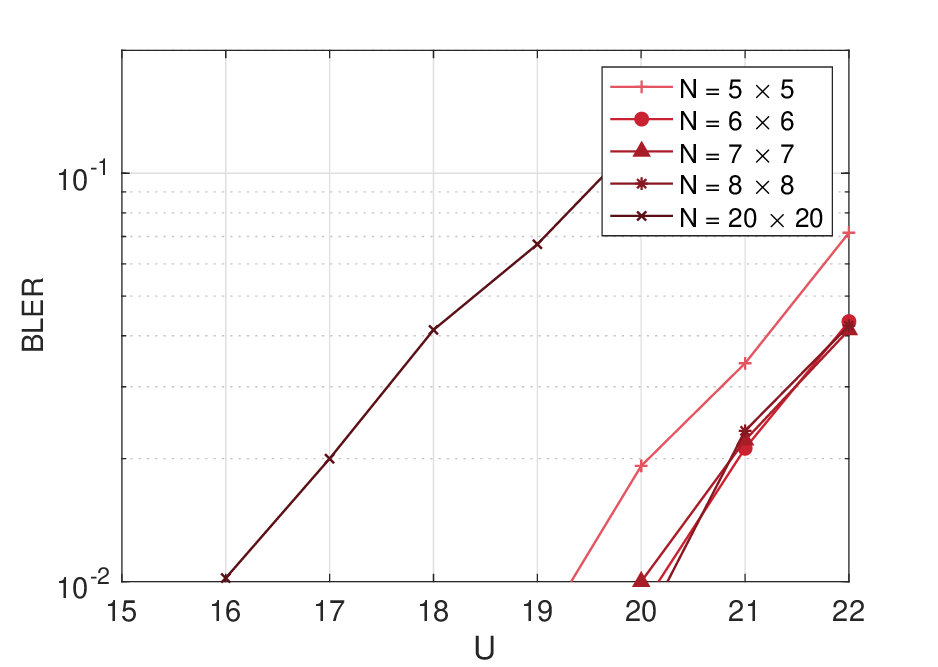}
\caption{BLER of MCS 7 against the number of UTs $U$, with $N_\text{RF} = 16$, $W = [2,2]$ over block fading channels.}\label{Fig:UvsBLER_diffN}
\vspace{3mm}
\end{figure}

Fig.~\ref{Fig:UvsBLER_diffN} presents the BLER of MCS 7 against the number of users, $U$, with $N_\text{RF} = 16$ and $W = [2,2]$. As can be seen, the error rate increases with $U$, since the interference will be more severe with more users. Besides, the FAS configurations of $N = 6\times 6$, $7\times 7$ or $8\times 8$ can accommodate more UTs while maintaining the same error rate, in contrast to the configurations of $N = 5\times 5$ or $20\times 20$. The minimum value of $N$ required to achieve the suboptimal performance is $N = 6\times 6$ for $W = [2,2]$ and $N_\text{RF} = 16$. Also, increasing the antenna ports to $20\times20$ results in great performance degradation. These observations further align with the semi-analytical evaluations of the achievable rates in Fig.~\ref{Fig:Ratevs1DN}. 

\vspace{-2mm}
\subsection{Practical Multiplexing Gain}\label{subsec:PMG}
Here, we conduct the link-level simulations of the suboptimal FAS configuration by Algorithm \ref{Alg:approxN} with $U = 2 N_\text{RF}$ and $C^{(N_1\times N_2)} = C_R$. Consequently, the numbers of antenna ports $N$ for different $W$ and $N_\text{RF}$ are given as in Table \ref{Tab:NforLLS}.

\begin{table}[t]
\centering
\caption{FAS configurations at UT of OFDM-FAMA} \label{Tab:NforLLS}
\begin{tabular}{c | c | c | c | c | c | c}
\hline
$W$ & \multicolumn{3}{c|}{$[2,2]$}& \multicolumn{3}{c}{$[5,5]$}\\ 
\hline
$N_\text{RF}$   & 2 & 4 & 16    & 2 & 4 & 16    \\ 
\hline\hline
$N$ &   $10\times 10$   & $8\times 8$   & $6\times 6$   &       $15\times 15$   & $12\times 12$   & $12\times 12$   \\ 
\hline
\end{tabular}
\end{table}

\begin{figure}[tbp]
\centering
\subfloat[]{\includegraphics[width = 0.9\linewidth]{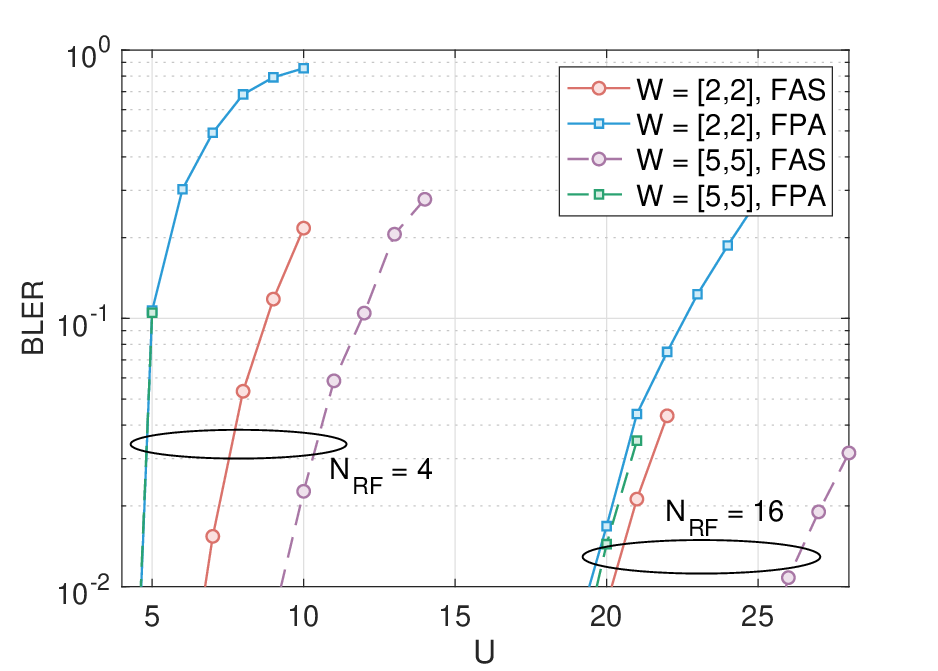}\label{SubFig:UvsBLER_BFC}}\\
\vspace{-4mm}
\subfloat[]{\includegraphics[width = 0.9\linewidth]{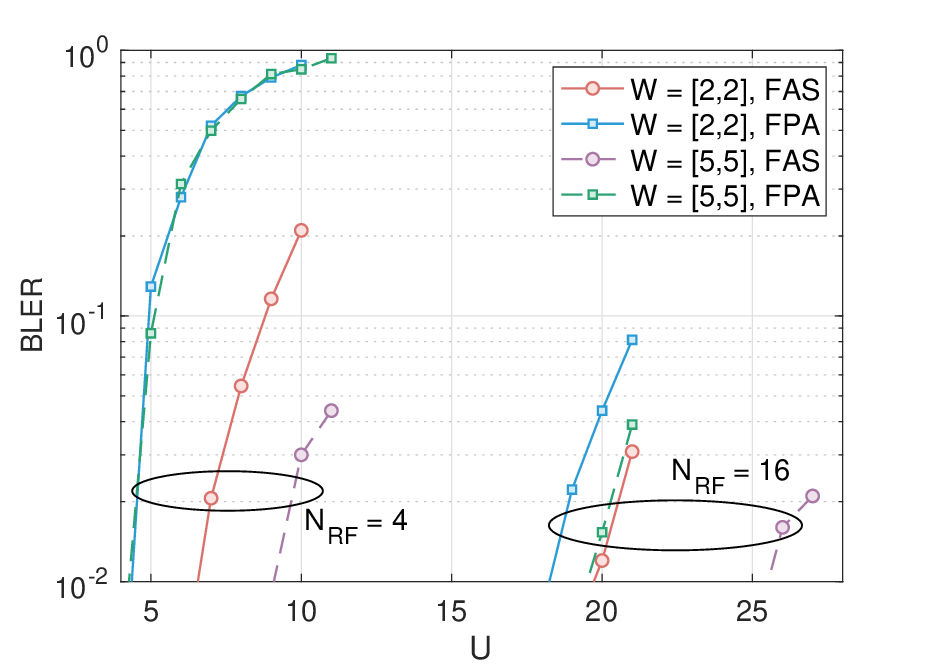}\label{SubFig:UvsBLER_TDL}}
\caption{The BLER performance of MCS7 against the number of UTs $U$, over (a) block fading, and (b) TDL-C channels.}\label{Fig:UvsBLER}
\end{figure}

The results in Fig.~\ref{Fig:UvsBLER} study the BLER of MCS 7 in relation to the number of users $U$, with different FAS configurations as detailed in Table \ref{Tab:NforLLS}. The results of the system with FPA at the UT are also included for comparison. In particular, we focus on identifying the maximum number of users when the BLER is below the threshold of $10^{-2}$. This number of users is considered as the practical multiplexing gain, which reflects the connectivity capability of the OFDM-FAMA system. It is evident that the practical multiplexing gain increases when a higher number of RF chains ($N_\text{RF}$) and a larger physical size of FAS ($W$). But the multiplexing gain remains relatively consistent as the physical size $W$ increases when FPA is utilized at the user. The performance enhancements of OFDM-FAMA, compared with the system with FPA, remains remarkable in the multipath channel environment, as shown in Fig.~\ref{Fig:UvsBLER}\ref{sub@SubFig:UvsBLER_TDL}. 

\begin{figure}
\centering
\subfloat[]{\includegraphics[width = 0.75\linewidth]{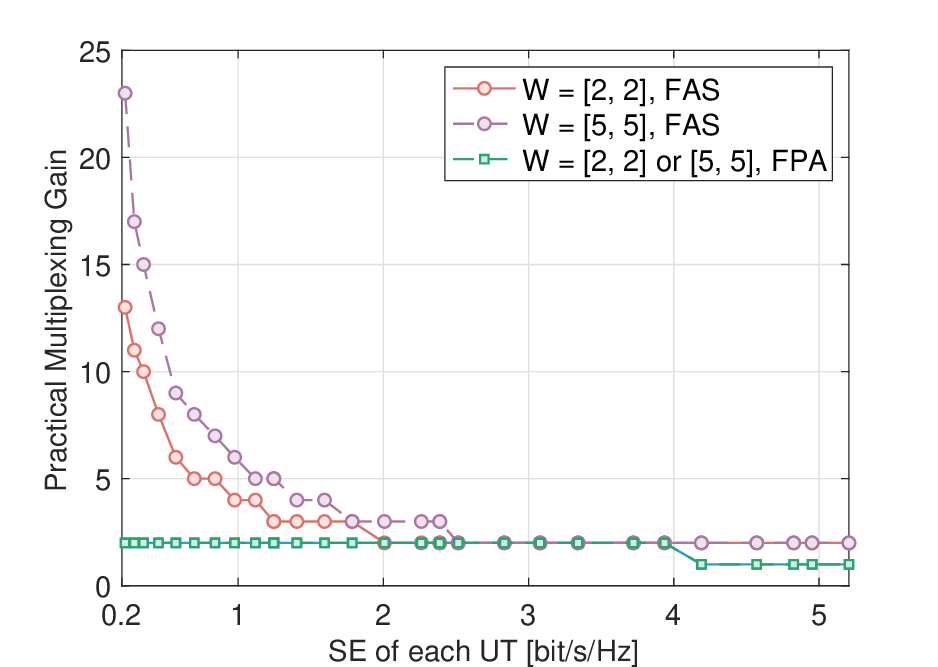}\label{SubFig:SEvsPMG_NRF2_BFL}}\\
\vspace{-4mm}
\subfloat[]{\includegraphics[width = 0.75\linewidth]{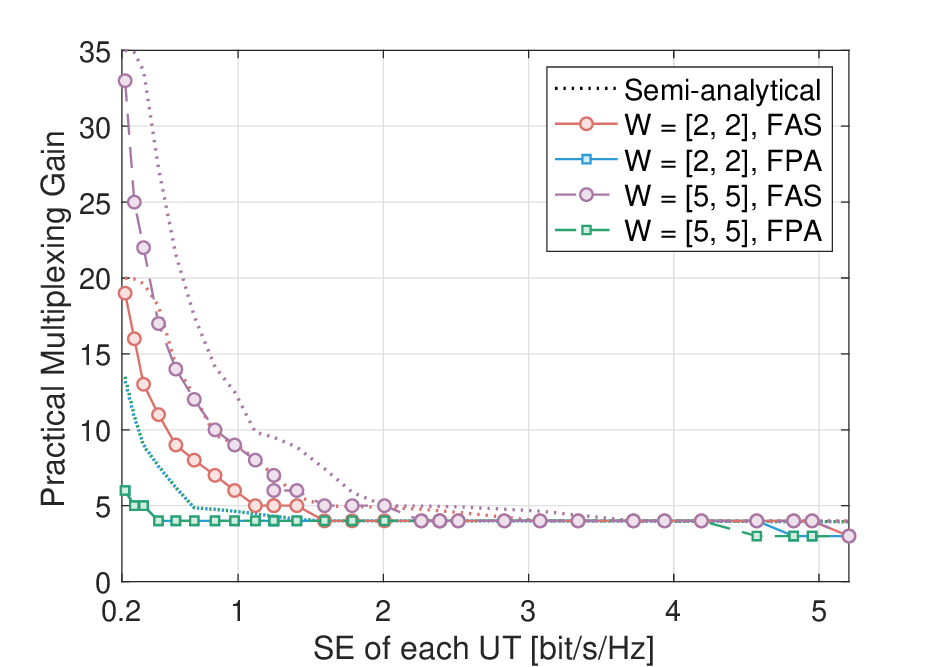}\label{SubFig:SEvsPMG_NRF4_BFL}}\\
\vspace{-4mm}
\subfloat[]{\includegraphics[width = 0.75\linewidth]{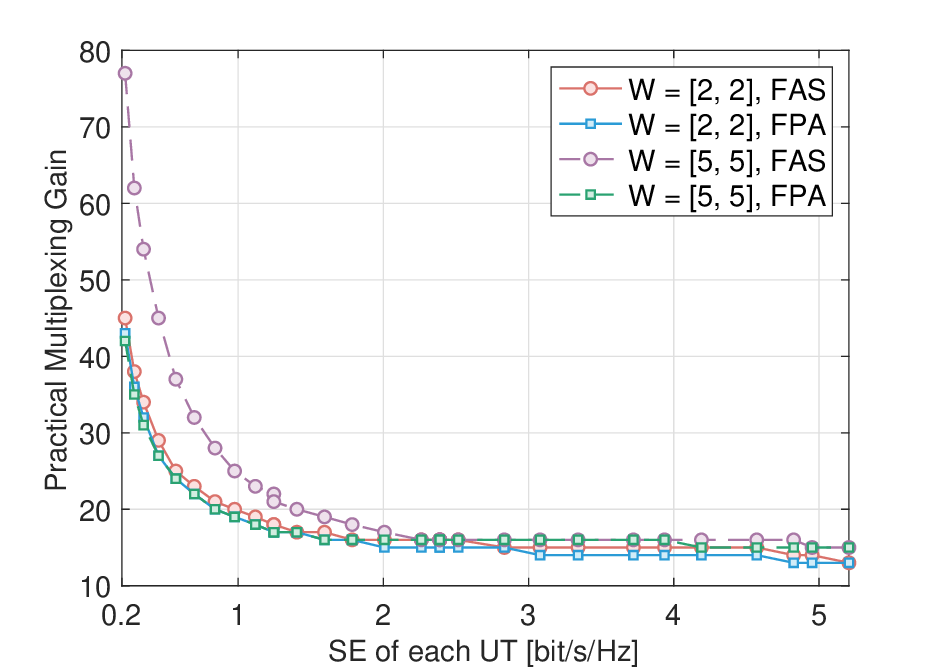}\label{SubFig:SEvsPMG_NRF16_BFL}}
\caption{Practical multiplexing gains of OFDM-FAMA against the SE over the block fading channel, with (a) $N_\text{RF} = 2$, (b) $N_\text{RF} = 4$, and (c) $N_\text{RF} = 16$.}\label{Fig:SEvsPMG_BFL}
\end{figure}

\begin{figure}
\centering
\subfloat[]{\includegraphics[width = 0.75\linewidth]{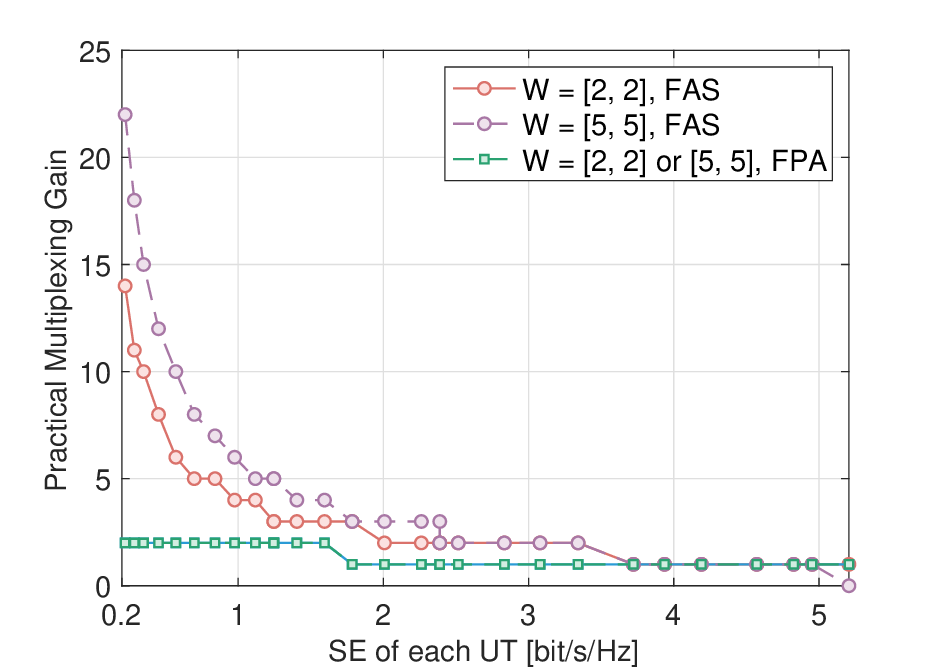}\label{SubFig:SEvsPMG_NRF2_TDL}}\\
\vspace{-4mm}
\subfloat[]{\includegraphics[width = 0.75\linewidth]{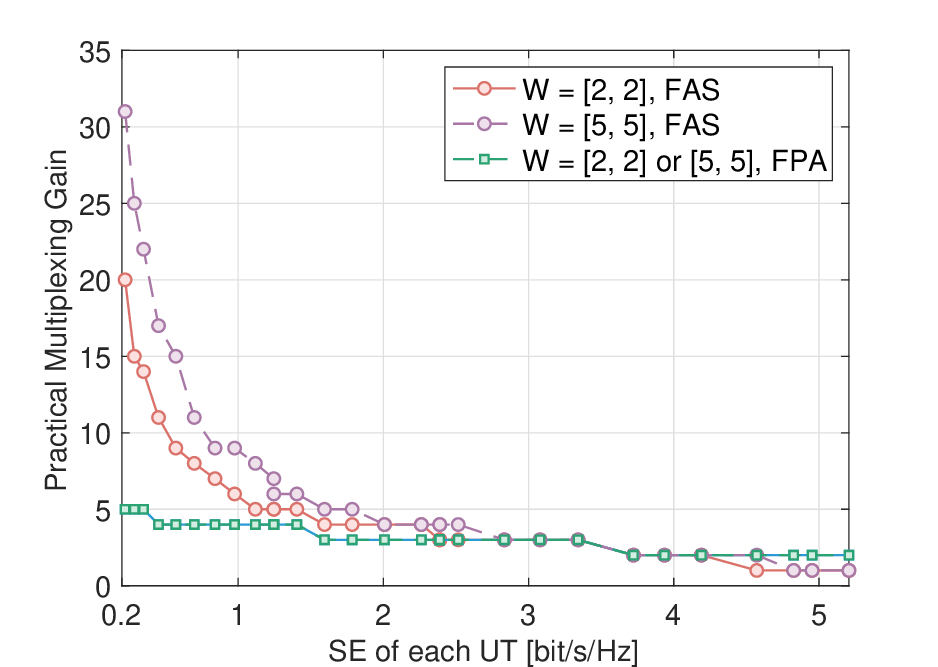}\label{SubFig:SEvsPMG_NRF4_TDL}}\\
\vspace{-4mm}
\subfloat[]{\includegraphics[width = 0.75\linewidth]{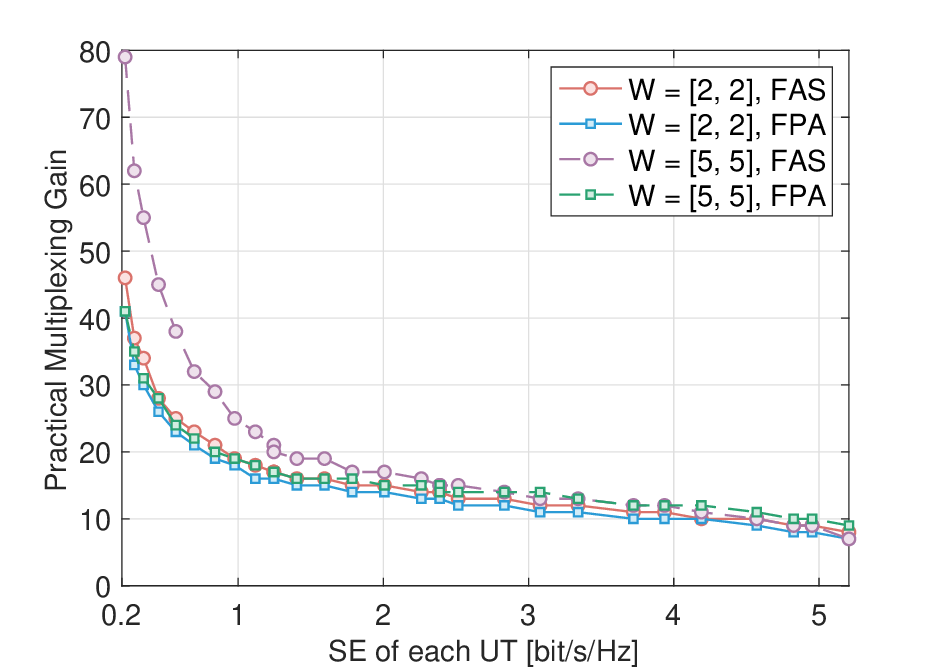}\label{SubFig:SEvsPMG_NRF16_TDL}}
\caption{Practical multiplexing gains of OFDM-FAMA against the SE over the TDL-C channel, with (a) $N_\text{RF} = 2$, (b) $N_\text{RF} = 4$, and (c) $N_\text{RF} = 16$.}\label{Fig:SEvsPMG_TDL}
\end{figure}

We conduct the link-level simulations for all the MCSs with SEs ranging from $0.2$ to $5.5$ bit/s/Hz over two channel models. The practical multiplexing gains for block fading and TDL-C are provided in Figs.~\ref{Fig:SEvsPMG_BFL} and \ref{Fig:SEvsPMG_TDL}, respectively. Here, the stationary scenario, with the Doppler frequency $f_d = 0$ Hz, is considered. When $N_\text{RF} = 4$, as depicted in Fig.~\ref{Fig:MGvsSE}\ref{sub@SubFig:SEvsPMG_NRF4_BFL}, the semi-analytical multiplexing gains in Section \ref{subsec:MG} are also given as the dotted lines for comparison. The semi-analytical multiplexing gain curves in this figure are derived from the outer envelopes of the multiplexing gain curves with different $U$ in Fig.~\ref{Fig:MGvsSE}. We observe that the practical multiplexing gain is slightly lower than the semi-analytical gain. This minor gap can be attributed to the rigorous criteria employed in the selection of the threshold, which is $\text{BLER} = 10^{-2}$ here.

The results in Figs.~\ref{Fig:SEvsPMG_BFL} and \ref{Fig:SEvsPMG_TDL} demonstrate that the OFDM-FAMA system offers substantial multiplexing gains when the SE of each UT is relatively low. A multiplexing gain of nearly $80$ is achievable at a rate of $0.2$ bit/s/Hz, when $N_\text{RF} = 16$ and $W = [5,5]$. The system can obtain a multiplexing gain exceeding $50$ UTs when using MCS 0-2. Notably, even with a limited number of RF chains at the receiver, the multiplexing gain can exceed $20$. This capability facilitates the connectivity of a large number of UTs in a spectrum-limited environment. However, it is observed that the multiplexing gain diminishes as SE increases, ultimately converging around $N_\text{RF}$. 

In comparison to the system with FPA, the OFDM-FAMA system demonstrates a greater multiplexing gain, particularly when the SE is relatively low. When the receiver at each UT is equipped with a limited number of RF chains, the multiplicative advantages become more pronounced. Specifically, the OFDM-FAMA system yields a ten-fold gain over the system with FPA when $N_\text{RF} = 2$, a five-fold gain when $N_\text{RF} = 4$, and a two-fold gain when $N_\text{RF} = 16$, with $W = [5,5]$ and adopting MCS 0. Conversely, the numerical value of the multiplexing gain increases as $N_\text{RF}$ rises. Specifically, the OFDM-FAMA system produces about $20$ more multiplexing gain compared to the system with FPA when $N_\text{RF} = 2$, about $25$ more when $N_\text{RF} = 4$, and roughly $40$ more when $N_\text{RF} = 16$, with $W = [5,5]$ and MCS 0 adopted. The significant multiplexing gains observed in low SE regions indicate that OFDM-FAMA could serve as a promising technology for mMTC, as mMTC traffic is characterized by low-rate transmissions.

Upon comparing the metrics with $W = [2,2]$ and $[5,5]$, we see that an increase in the antenna size does not yield a significant gain in the system with FPA. However, in OFDM-FAMA, a higher practical multiplexing gain is observed with a larger physical size when SE is low. When $W = [2,2]$ and $N_\text{RF} = 16$, the performance of OFDM-FAMA closely resembles that of the FPA-based system. This aligns with the semi-analytical results in Fig.~\ref{Fig:Ratevs1DN}. The reason for this similarity is that the FPA MIMO system with $N = 4\times 4$ receive antennas over a size of $W = 2\lambda \times 2\lambda$ almost approaches the saturation point, thereby exhibiting a relatively high connectivity capability.

Comparing the two channel models, we can first state that their performances are similar at low SE. When SE is elevated, the multiplexing gain over the TDL channel is slightly lower. Also, the range of SE values in which OFDM-FAMA shows a superior multiplexing gain compared to the system using FPA is more extensive within the TDL-C channel.

\vspace{-4mm}
\subsection{Mobility Evaluation}\label{subsec:Mob}
In order to evaluate the mobility, the TDL-C channel model is used, with the maximum Doppler shift set as $30$ Hz for the low-speed scenario and $300$ Hz for the high-speed scenario. The practical multiplexing gains for different mobility scenarios are presented in the Tables \ref{Tab:PMG_fD0Hz}--\ref{Tab:PMG_fD300Hz}.

\begin{table*}
\centering
\caption{Practical multiplexing gains of the OFDM-FAMA system in the stationary scenario when $f_d = 0$ Hz}\label{Tab:PMG_fD0Hz}
\begin{tabular}{c|c|c|c|c|c|c|c|c|c|c|c|c|c|c|c|c}
\hline
MCS & Mod.& Target& \multirow{3}{*}{TBS}& SE & \multicolumn{6}{c|}{$W = [2,2]$} & \multicolumn{6}{c}{$W = [5,5]$}\\ 
\cline{6-17}
Index & Order& CR & & per UT & \multicolumn{2}{c|}{$N_\text{RF} = 16$} & \multicolumn{2}{c|}{$N_\text{RF} = 4$} & \multicolumn{2}{c|}{$N_\text{RF} = 2$} & \multicolumn{2}{c|}{$N_\text{RF} = 16$} & \multicolumn{2}{c|}{$N_\text{RF} = 4$} & \multicolumn{2}{c}{$N_\text{RF} = 2$}\\ 
\cline{6-17}
$I_\text{MCS}$& $Q_m$& $\times1024$ & & [bit/s/Hz] & FAS & FPA & FAS & FPA & FAS & FPA & FAS & FPA & FAS & FPA & FAS & FPA \\ \hline\hline
        0&  2&  120&    224&    0.2222& 46& 41& 20& 5& 15& 2& 79& 41& 31& 5& 22& 2 \\ \hline
        1&  2&  157&    288&    0.2857& 37&	33& 15& 5& 11& 2& 62& 35& 25& 5& 18& 2  \\ \hline
        2&  2&  193&    352&    0.3492&	34&	30&	14&	5&	10&	2&	55&	31&	22&	5&	15&	2   \\ \hline
        3&  2&  251&	456&    0.4524&	28&	26&	11&	4&	8&	2&	45&	28&	17&	4&	12&	2   \\ \hline
        4&  2&  308&    576&    0.5714&	25&	23&	9&	4&	6&	2&	38&	24&	15&	4&	10&	2   \\ \hline
        5&  2&  379&    704&    0.6984&	23&	21&	8&	4&	5&	2&	32&	22&	11&	4&	8&	2   \\ \hline
        6&  2&  449&	848&    0.8413&	21&	19&	7&	4&	5&	2&	29&	20&	9&	4&	7&	2   \\ \hline
        7&  2&  526&	984&    0.9762&	19&	18&	6&	4&	4&	2&	25&	19&	9&	4&	6&	2   \\ \hline
        8&  2&  602&	1128&   1.1190&	18&	16&	5&	4&	4&	2&	23&	18&	8&	4&	5&	2   \\ \hline
        9&  2&  679&	1256&   1.2460&	17&	16&	5&	4&	3&	2&	21&	17&	7&	4&	5&	2   \\ \hline
        10& 4&  340&	1256&   1.2460&	17&	16&	5&	4&	3&	2&	20&	17&	6&	4&	5&	2   \\ \hline
        11& 4&  378&	1416&   1.4048&	16&	15&	5&	4&	3&	2&	19&	16&	6&	4&	4&	2   \\ \hline
        12& 4&  434&	1608&   1.5952&	16&	15&	4&	3&	3&	2&	19&	16&	5&	3&	4&	2   \\ \hline
        13& 4&  490&	1800&   1.7857&	15&	14&	4&	3&	3&	1&	17&	16&	5&	3&	3&	1   \\ \hline
        14& 4&  553&	2024&   2.0079&	15&	14&	4&	3&	2&	1&	17&	15&	4&	3&	3&	1   \\ \hline
        15& 4&  616&	2280&   2.2619&	14&	13&	4&	3&	2&	1&	16&	15&	4&	3&	3&	1   \\ \hline
        16& 4&  658&	2408&   2.3889&	14&	13&	\cellcolor{gray}3&	\cellcolor{gray}3&	2&	1&	\cellcolor{gray}15&	\cellcolor{gray}15&	4&	3&	3&	1   \\ \hline
        17& 6&  438&	2408&   2.3889&	14&	13&	\cellcolor{gray}3&	\cellcolor{gray}3&	2&	1&	15&	14&	4&	3&	2&	1   \\ \hline
        18& 6&  466&	2536&   2.5159&	13&	12&	\cellcolor{gray}3&	\cellcolor{gray}3&	2&	1&	15&	14&	4&	3&	2&	1   \\ \hline
        19& 6&  517&	2856&   2.8333&	13&	12&	\cellcolor{gray}3&	\cellcolor{gray}3&	2&	1&	\cellcolor{gray}14&	\cellcolor{gray}14&	\cellcolor{gray}3&	\cellcolor{gray}3&	2&	1   \\ \hline
        20& 6&  567&	3104&   3.0794&	12&	11&	\cellcolor{gray}3&	\cellcolor{gray}3&	2&	1&	\cellcolor{gray}13&	\cellcolor{gray}14&	\cellcolor{gray}3&	\cellcolor{gray}3&	2&	1   \\ \hline
        21& 6&  616&	3368&   3.3413&	12&	11&		\cellcolor{gray}3&	\cellcolor{gray}3&	2&	1&	\cellcolor{gray}13&	\cellcolor{gray}13&	\cellcolor{gray}3&	\cellcolor{gray}3&	2&	1   \\ \hline
        22& 6&  666&	3752&   3.7222&	11&	10&		\cellcolor{gray}2&	\cellcolor{gray}2&	\cellcolor{gray}1&	\cellcolor{gray}1&	\cellcolor{gray}12&	\cellcolor{gray}12&	\cellcolor{gray}2&	\cellcolor{gray}2&	\cellcolor{gray}1&	\cellcolor{gray}1   \\ \hline
        23& 6&  719&	3968&   3.9365&	11&	10&	\cellcolor{gray}2&	\cellcolor{gray}2&	\cellcolor{gray}1&	\cellcolor{gray}1&	\cellcolor{gray}12&	\cellcolor{gray}12&	\cellcolor{gray}2&	\cellcolor{gray}2&	\cellcolor{gray}1&	\cellcolor{gray}1   \\ \hline
        24& 6&  772&	4224&   4.1905&	\cellcolor{gray}10&	\cellcolor{gray}10&	\cellcolor{gray}2&	\cellcolor{gray}2&	\cellcolor{gray}1&	\cellcolor{gray}1&	\cellcolor{gray}11&	\cellcolor{gray}12&	\cellcolor{gray}2&	\cellcolor{gray}2&	\cellcolor{gray}1&	\cellcolor{gray}1   \\ \hline
        25& 6&  822&	4608&   4.5714&	10&	9&	\cellcolor{gray}1&	\cellcolor{gray}2&	\cellcolor{gray}1&	\cellcolor{gray}1&	\cellcolor{gray}10&	\cellcolor{gray}11&	\cellcolor{gray}2&	\cellcolor{gray}2&	\cellcolor{gray}1&	\cellcolor{gray}1   \\ \hline
        26& 6&  873&	4864&   4.8254&	9&	8&	\cellcolor{gray}1&	\cellcolor{gray}2&	\cellcolor{gray}1&	\cellcolor{gray}1&	\cellcolor{gray}9&	\cellcolor{gray}10&	\cellcolor{gray}1&	\cellcolor{gray}2&	\cellcolor{gray}1&	\cellcolor{gray}1   \\ \hline
        27& 6&  910&	4992&   4.9524&	9&	8&	\cellcolor{gray}1&	\cellcolor{gray}2&	\cellcolor{gray}1&	\cellcolor{gray}1&	\cellcolor{gray}9&	\cellcolor{gray}10&	\cellcolor{gray}1&	\cellcolor{gray}2&	\cellcolor{gray}1&	\cellcolor{gray}1   \\ \hline
        28& 6&  948&	5248&   5.2063&	8&	7& \cellcolor{gray}1&	 \cellcolor{gray}2& \cellcolor{gray}1&	\cellcolor{gray}1& \cellcolor{gray}7&	\cellcolor{gray}9& \cellcolor{gray}1& \cellcolor{gray}2& \cellcolor{gray}0& 	\cellcolor{gray}1   \\ \hline
\end{tabular}
\begin{tablenotes}
\item Note that the grey-shaded cells indicate the cases where FAS does not outperform FPA.
\end{tablenotes}
\vspace{-3mm}
\end{table*}

\begin{table*}
\centering
\caption{Practical multiplexing gains of the OFDM-FAMA system in the low-speed scenario when $f_d = 30$ Hz}\label{Tab:PMG_fD30Hz}
\begin{tabular}{c|c|c|c|c|c|c|c|c|c|c|c|c|c|c|c|c}
\hline
MCS & Mod.& Target& \multirow{3}{*}{TBS}& SE & \multicolumn{6}{c|}{$W = [2,2]$} & \multicolumn{6}{c}{$W = [5,5]$}\\ 
\cline{6-17}
Index & Order& CR & & per UT & \multicolumn{2}{c|}{$N_\text{RF} = 16$} & \multicolumn{2}{c|}{$N_\text{RF} = 4$} & \multicolumn{2}{c|}{$N_\text{RF} = 2$} & \multicolumn{2}{c|}{$N_\text{RF} = 16$} & \multicolumn{2}{c|}{$N_\text{RF} = 4$} & \multicolumn{2}{c}{$N_\text{RF} = 2$}\\ \cline{6-17}
        $I_\text{MCS}$& $Q_m$& $\times1024$ & & [bit/s/Hz] & FAS & FPA & FAS & FPA & FAS & FPA & FAS & FPA & FAS & FPA & FAS & FPA \\ \hline\hline
        0&	2&	120&	224&	0.2222&	46&	41&	20&	6&	14&	2&	80&	42&		32&6&	21&	2\\ \hline
        1&	2&	157&	288&	0.2857&	39&	34&	16&	5&	11&	2&	63&	35&		25&5&	17&	2\\ \hline
        2&	2&	193&	352&	0.3492&	35&	30&	14&	5&	10&	2&	53&	31&		21&5&	17&	2\\ \hline
        3&	2&	251&	456&	0.4524&	29&	26&	11&	4&	8&	2&	46&	28&     17& 4&	11&	2\\ \hline
        4&	2&	308&	576&	0.5714&	25&	23&	9&	4&	6&	2&	36&	24&		14&4&	9&	2\\ \hline
        5&	2&	379&	704&	0.6984&	22&	20&	8&	4&	5&	2&	33&	22&		11&4&	8&	2\\ \hline
        6&	2&	449&	848&	0.8413&	20&	18&	7&	4&	5&	2&	28&	20&		10&4&	7&	2\\ \hline
        7&	2&	526&	984&	0.9762&	19&	18&	6&	4&	4&	2&	26&	19&		8&4&	6&	2\\ \hline
        8&	2&	602&	1128&	1.1190&	18&	17&	6&	4&	4&	2&	23&	18&	8&	4&	5&	2\\ \hline
        9&	2&	679&	1256&	1.2460&	17&	16&	5&	4&	3&	2&	21&	17&	7&	3&	5&	2\\ \hline
        10&	4&	340&	1256&	1.2460&	17&	16&	5&	4&	3&	2&	21&	17&	7&	3&	5&	1\\ \hline
        11&	4&	378&	1416&	1.4048&	16&	15&	4&	3&	3&	1&	19&	16&	6&	3&	4&	1\\ \hline
        12&	4&	434&	1608&	1.5952&	16&	14&	4&	3&	3&	1&	18&	16&	5&	3&	4&	1\\ \hline
        13&	4&	490&	1800&	1.7857&	15&	14&	4&	3&	3&	1&	17&	15&	5&	3&	3&	1\\ \hline
        14&	4&	553&	2024&	2.0079&	14&	13&	4&	3&	2&	1&	16&	15&	5&	3&	3&	1\\ \hline
        15&	4&	616&	2280&	2.2619&	14&	13&	\cellcolor{gray}3&	\cellcolor{gray}3&	2&	1&	16&	14&	4&	3&	3&	1\\ \hline
        16&	4&	658&	2408&	2.3889&	14&	13&	\cellcolor{gray}3&	\cellcolor{gray}3&	2&	1&	15&	14&	4&	3&	3&	1\\ \hline
        17&	6&	438&	2408&	2.3889&	13&	12&	\cellcolor{gray}3&	\cellcolor{gray}3&	2&	1&	15&	14&	4&	3&	2&	1\\ \hline
        18&	6&	466&	2536&	2.5159&	13&	12&	\cellcolor{gray}3&	\cellcolor{gray}3&	2&	1&	\cellcolor{gray}14&	\cellcolor{gray}14&	4&	3&	2&	1\\ \hline
        19&	6&	517&	2856&	2.8333&	\cellcolor{gray}12&	\cellcolor{gray}12&	\cellcolor{gray}3&	\cellcolor{gray}3&	2&	1&	14&	13&	\cellcolor{gray}3&	\cellcolor{gray}3&	2&	1\\ \hline
        20&	6&	567&	3104&	3.0794&	12&	11&	\cellcolor{gray}3&	\cellcolor{gray}3&	2&	1&	\cellcolor{gray}13&	\cellcolor{gray}13&	\cellcolor{gray}3&	\cellcolor{gray}3&	2&	1\\ \hline
        21&	6&	616&	3368&	3.3413&	12&	10&	3&	2&	\cellcolor{gray}1&	\cellcolor{gray}1&	13&	12&	\cellcolor{gray}3&	\cellcolor{gray}3&	2&	1\\ \hline
        22&	6&	666&	3752&	3.7222&	\cellcolor{gray}11&	\cellcolor{gray}10&	\cellcolor{gray}2&	\cellcolor{gray}2&	\cellcolor{gray}1&	\cellcolor{gray}1&	\cellcolor{gray}12&	\cellcolor{gray}12&	\cellcolor{gray}2&	\cellcolor{gray}2&	\cellcolor{gray}1&	\cellcolor{gray}1\\ \hline
        23&	6&	719&	3968&	3.9365&	\cellcolor{gray}10&	\cellcolor{gray}10&	\cellcolor{gray}2&	\cellcolor{gray}2&	\cellcolor{gray}1&	\cellcolor{gray}1&	\cellcolor{gray}11&	\cellcolor{gray}12&	\cellcolor{gray}2&	\cellcolor{gray}2&	\cellcolor{gray}1&	\cellcolor{gray}1\\ \hline
        24&	6&	772&	4224&	4.1905&	10&	9&	\cellcolor{gray}2&	\cellcolor{gray}2&	\cellcolor{gray}1&	\cellcolor{gray}1&	\cellcolor{gray}11&	\cellcolor{gray}11&	\cellcolor{gray}2&	\cellcolor{gray}2&	\cellcolor{gray}1&	\cellcolor{gray}1\\ \hline
        25&	6&	822&	4608&	4.5714&	9&	8&	\cellcolor{gray}2&	\cellcolor{gray}2&	\cellcolor{gray}1&	\cellcolor{gray}1&	\cellcolor{gray}10&	\cellcolor{gray}11&	\cellcolor{gray}2&	\cellcolor{gray}2&	\cellcolor{gray}1&	\cellcolor{gray}1\\ \hline
        26&	6&	873&	4864&	4.8254&	\cellcolor{gray}8&	\cellcolor{gray}8&	\cellcolor{gray}1&	\cellcolor{gray}2&	\cellcolor{gray}1&	\cellcolor{gray}1&	\cellcolor{gray}9&	\cellcolor{gray}9&	\cellcolor{gray}1&	\cellcolor{gray}2&	\cellcolor{gray}1&	\cellcolor{gray}1\\ \hline
        27&	6&	910&	4992&	4.9524&	8&	7&	\cellcolor{gray}1&	\cellcolor{gray}2&	\cellcolor{gray}1&	\cellcolor{gray}1&	\cellcolor{gray}8&	\cellcolor{gray}9&	\cellcolor{gray}1&	\cellcolor{gray}2&	\cellcolor{gray}1&	\cellcolor{gray}1\\ \hline
        28&	6&	948&	5248&	5.2063&	\cellcolor{gray}7&	\cellcolor{gray}7&	\cellcolor{gray}1&	\cellcolor{gray}2&	\cellcolor{gray}1&	\cellcolor{gray}1&	\cellcolor{gray}7&	\cellcolor{gray}8&	\cellcolor{gray}1&	\cellcolor{gray}2&	\cellcolor{gray}0&	\cellcolor{gray}1\\ \hline
\end{tabular}
\begin{tablenotes}
\item Note that the grey-shaded cells indicate the cases where FAS does not outperform FPA.
\end{tablenotes}
\vspace{-3mm}
\end{table*}

\begin{table*}
\centering
\caption{Practical multiplexing gains of the OFDM-FAMA system in the high-speed scenario when $f_d = 300$ Hz}\label{Tab:PMG_fD300Hz}
\begin{tabular}{c|c|c|c|c|c|c|c|c|c|c|c|c|c|c|c|c}
\hline
MCS & Mod.& Target& \multirow{3}{*}{TBS}& SE & \multicolumn{6}{c|}{$W = [2,2]$} & \multicolumn{6}{c}{$W = [5,5]$}\\ \cline{6-17}
Index & Order& CR & & per UT & \multicolumn{2}{c|}{$N_\text{RF} = 16$} & \multicolumn{2}{c|}{$N_\text{RF} = 4$} & \multicolumn{2}{c|}{$N_\text{RF} = 2$} & \multicolumn{2}{c|}{$N_\text{RF} = 16$} & \multicolumn{2}{c|}{$N_\text{RF} = 4$} & \multicolumn{2}{c}{$N_\text{RF} = 2$}\\ \cline{6-17}
$I_\text{MCS}$& $Q_m$& $\times1024$ & & [bit/s/Hz] & FAS & FPA & FAS & FPA & FAS & FPA & FAS & FPA & FAS & FPA & FAS & FPA \\ \hline\hline
        0&	2&	120&	224&	0.2222&	44&	41&	18&	6&	11&	2&	80&	41&	31&	6&	18&	2\\ \hline
        1&	2&	157&	288&	0.2857&	37&	33&	14&	5&	9&	2&	56&	35&	22&	5&	14&	2\\ \hline
        2&	2&	193&	352&	0.3492&	32&	29&	12&	4&	8&	1&	52&	30&	19&	4&	12&	1\\ \hline
        3&	2&	251&	456&	0.4524&	28&	24&	10&	4&	6&	1&	42&	25&	16&	3&	10&	1\\ \hline
        4&	2&	308&	576&	0.5714&	23&	21&	8&	3&	5&	1&	35&	23&	12&	3&	8&	1\\ \hline
        5&	2&	379&	704&	0.6984&	20&	18&	7&	3&	4&	1&	29&	19&	10&	3&	6&	1\\ \hline
        6&	2&	449&	848&	0.8413&	18&	16&	6&	3&	4&	1&	24&	17&	9&	3&	5&	1\\ \hline
        7&	2&	526&	984&	0.9762&	16&	14&	5&	3&	3&	1&	21&	15&	7&	3&	4&	1\\ \hline
        8&	2&	602&	1128&	1.1190&	14&	13&	5&	3&	3&	1&	19&	14&	7&	3&	4&	1\\ \hline
        9&	2&	679&	1256&	1.2460&	13&	12&	4&	2&	3&	1&	18&	13&	6&	2&	4&	1\\ \hline
        10&	4&	340&	1256&	1.2460&	13&	12&	4&	2&	3&	1&	17&	13&	5&	2&	4&	1\\ \hline
        11&	4&	378&	1416&	1.4048&	12&	11&	4&	2&	2&	1&	16&	12&	5&	2&	3&	1\\ \hline
        12&	4&	434&	1608&	1.5952&	11&	10&	3&	2&	2&	1&	14&	11&	4&	2&	3&	1\\ \hline
        13&	4&	490&	1800&	1.7857&	\cellcolor{gray}10&	\cellcolor{gray}10&	3&	2&	2&	1&	13&	11&	4&	2&	2&	1\\ \hline
        14&	4&	553&	2024&	2.0079&	\cellcolor{gray}9&	\cellcolor{gray}9&	3&	2&	2&	1&	12&	10&	3&	2&	2&	1\\ \hline
        15&	4&	616&	2280&	2.2619&	9&	8&	3&	2&	2&	1&	11&	9&	3&	2&	2&	1\\ \hline
        16&	4&	658&	2408&	2.3889&	9&	8&	\cellcolor{gray}2&	\cellcolor{gray}2&	2&	1&	10&	9&	3&	2&	2&	1\\ \hline
        17&	6&	438&	2408&	2.3889&	\cellcolor{gray}8&	\cellcolor{gray}8&	\cellcolor{gray}2&	\cellcolor{gray}2&	\cellcolor{gray}1&	\cellcolor{gray}1&	10&	9&	3&	2&	2&	1\\ \hline
        18&	6&	466&	2536&	2.5159&	\cellcolor{gray}8&	\cellcolor{gray}8&	\cellcolor{gray}2&	\cellcolor{gray}2&	\cellcolor{gray}1&	\cellcolor{gray}1&	10&	9&	3&	2&	2&	1\\ \hline
        19&	6&	517&	2856&	2.8333&	\cellcolor{gray}7&	\cellcolor{gray}7&	\cellcolor{gray}2&	\cellcolor{gray}2&	\cellcolor{gray}1&	\cellcolor{gray}1&	9&	8&	\cellcolor{gray}2&	\cellcolor{gray}2&	2&	1\\ \hline
        20&	6&	567&	3104&	3.0794&	\cellcolor{gray}7&	\cellcolor{gray}7&	\cellcolor{gray}2&	\cellcolor{gray}2&	\cellcolor{gray}1&	\cellcolor{gray}1&	\cellcolor{gray}8&	\cellcolor{gray}8&	\cellcolor{gray}2&	\cellcolor{gray}2&	\cellcolor{gray}1&	\cellcolor{gray}1\\ \hline
        21&	6&	616&	3368&	3.3413&	7&	6&	\cellcolor{gray}2&	\cellcolor{gray}2&	\cellcolor{gray}1&	\cellcolor{gray}1&	\cellcolor{gray}8&	\cellcolor{gray}8&	\cellcolor{gray}2&	\cellcolor{gray}2&	\cellcolor{gray}1&	\cellcolor{gray}1\\ \hline
        22&	6&	666&	3752&	3.7222&	\cellcolor{gray}6&	\cellcolor{gray}6&	\cellcolor{gray}2&	\cellcolor{gray}2&	\cellcolor{gray}1&	\cellcolor{gray}1&	\cellcolor{gray}7&	\cellcolor{gray}7&	\cellcolor{gray}2&	\cellcolor{gray}2&	\cellcolor{gray}1&	\cellcolor{gray}1\\ \hline
        23&	6&	719&	3968&	3.9365&	\cellcolor{gray}6&	\cellcolor{gray}6&	2&	1&	\cellcolor{gray}1&	\cellcolor{gray}1&	\cellcolor{gray}7&	\cellcolor{gray}7&	\cellcolor{gray}2&	\cellcolor{gray}2&	\cellcolor{gray}1&	\cellcolor{gray}1\\ \hline
        24&	6&	772&	4224&	4.1905&	6&	5&	\cellcolor{gray}1&	\cellcolor{gray}1&	\cellcolor{gray}1&	\cellcolor{gray}1&	\cellcolor{gray}6&	\cellcolor{gray}7&	\cellcolor{gray}1&	\cellcolor{gray}1&	\cellcolor{gray}1&	\cellcolor{gray}1\\ \hline
        25&	6&	822&	4608&	4.5714&	\cellcolor{gray}5&	\cellcolor{gray}5&	\cellcolor{gray}1&	\cellcolor{gray}1&	\cellcolor{gray}1&	\cellcolor{gray}1&	\cellcolor{gray}6&	\cellcolor{gray}6&	\cellcolor{gray}1&	\cellcolor{gray}1&	\cellcolor{gray}1&	\cellcolor{gray}1\\ \hline
        26&	6&	873&	4864&	4.8254&	\cellcolor{gray}5&	\cellcolor{gray}5&	\cellcolor{gray}1&	\cellcolor{gray}1&	\cellcolor{gray}1&	\cellcolor{gray}1&	\cellcolor{gray}5&	\cellcolor{gray}6&	\cellcolor{gray}1&	\cellcolor{gray}1&	\cellcolor{gray}1&	\cellcolor{gray}1\\ \hline
        27&	6&	910&	4992&	4.9524&	\cellcolor{gray}5&	\cellcolor{gray}5&	\cellcolor{gray}1&	\cellcolor{gray}1&	\cellcolor{gray}1&	\cellcolor{gray}1&	\cellcolor{gray}5&	\cellcolor{gray}6&	\cellcolor{gray}1&	\cellcolor{gray}1&	\cellcolor{gray}0&	\cellcolor{gray}1\\ \hline
        28&	6&	948&	5248&	5.2063&	\cellcolor{gray}4&	\cellcolor{gray}4&	\cellcolor{gray}1&	\cellcolor{gray}1&	\cellcolor{gray}0&	\cellcolor{gray}0&	\cellcolor{gray}4&	\cellcolor{gray}5&	\cellcolor{gray}1&	\cellcolor{gray}1&	\cellcolor{gray}0&	\cellcolor{gray}0\\ \hline
\end{tabular}
\begin{tablenotes}
\item Note that the grey-shaded cells indicate the cases where FAS does not outperform FPA.
\end{tablenotes}
\vspace{-3mm}
\end{table*}

The results in the tables indicate that OFDM-FAMA maintains high multiplexing gains with robust MCS at low SE, even in the high-speed scenario. For example, the multiplexing gain of OFDM-FAMA, with UTs with $N_\text{RF} = 16$ RF chains and a FAS of ($N = 12\times 12$, $W = [5,5]$), remains approximately $80$ if MCS 0 is used. Furthermore, OFDM-FAMA can support $25$ UTs in the stationary scenario, $26$ UTs in the low-speed scenario, and $21$ UTs in the high-speed scenario, when MCS 7 with $\text{SE} = 1.0$ bit/s/Hz is applied. Besides, OFDM-FAMA continues to provide gains in the mobility scenarios compared to the FPA-based system, particularly at low SE. Nonetheless, as the SE increases and high-order modulation is used, the multiplexing gain diminishes in the stationary scenario, and further declines with increasing Doppler frequency.

\vspace{-2mm}
\subsection{Performance of Training Stage}\label{subsec:PerformanceTS}

\begin{figure}
\centering
\includegraphics[width = \linewidth]{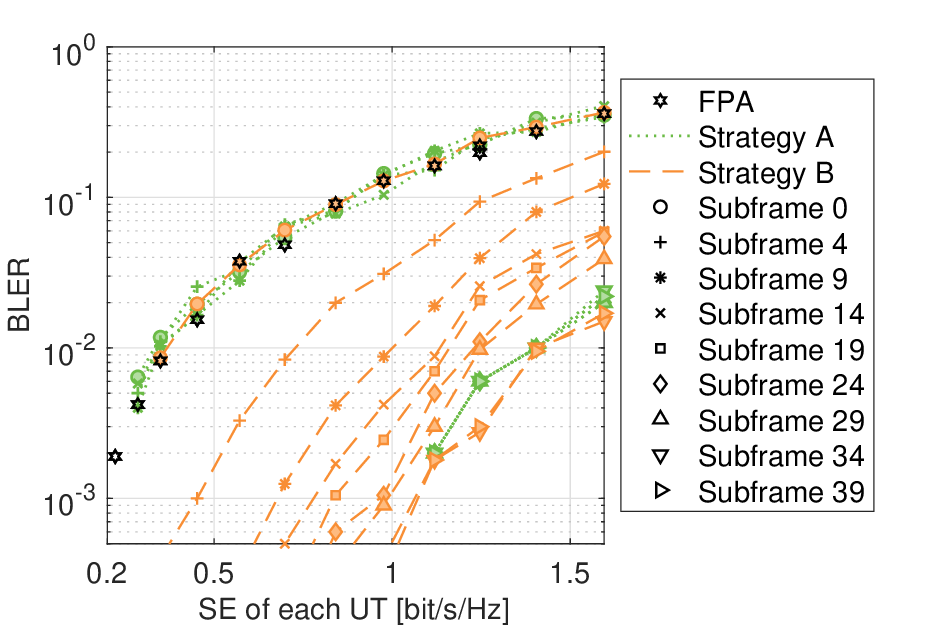}
\caption{BLER performances against the SE of OFDM-FAMA for $40$ continuous subframes over TDL-C channels, when $U = 5$, $N_\text{RF} = 4$, $N = 8\times 8$, and $W = [2,2]$.}\label{Fig:SEvsBLER_TS}
\end{figure}

Here we evaluate the performance during the training stage with parameters: $N_\text{RF} = 4$, $N = 8\times 8$, and $W = [2,2]$. The number of subframes designated for training is $N_\text{subframe}^\text{TSA} = 16$ for \emph{Strategy A}, and $N_\text{subframe}^\text{TSB} = 31$ for \emph{Strategy B}. We simulate $40$ continuous subframes with $U = 5$ UTs in the OFDM-FAMA system. The BLERs of these continuous subframes are illustrated in Fig.~\ref{Fig:SEvsBLER_TS}, against the SE per UT under different MCSs. As we can observe, the BLER during the training stage is inferior to that of the running stage, indicating that the system needs to operate at a low rate during the training stage, and the rate could subsequently be increased during the running stage. The curves of \emph{Strategy A} are categorized into two distinct clusters. The first cluster pertains to the subframes within the training stage, i.e., $n_\text{subframe} < N_\text{subframe}^\text{TSA}$, where their performance aligns with that of the FPA-based system. The second cluster corresponds to the subframes in the running stage, i.e., $n_\text{subframe} \geq N_\text{subframe}^\text{TSA}$, where the BLER performance is significantly enhanced compared to the training stage. In the case of \emph{Strategy B}, the performance of the first subframe ($n_\text{subframe}=0$) is the same as that of the system with FPA, and continuous improvement in the performance is noted as $n_\text{subframe}$ increases. This occurs because \emph{Strategy B} selects half of the RF chains from the known good antenna ports. 

\vspace{-2mm}
\section{Conclusion}\label{sec:conclusion}
In this paper, we proposed the OFDM-FAMA system, using the 5G NR numerology and channel coding. The average SINR of a subframe serves as a key metric for selecting the ports for FAMA, and we proposed the port selector with two training strategies. IRC equalization was employed to further mitigate the interference at the receiver. We derived and conducted the analysis of the outage rate, channel capacity, and cutoff rate, and proposed an algorithm to find the suboptimal configuration of FAS at the UT based on these rates. Link-level results in this paper confirmed the suboptimal $N^*$ identified through the proposed algorithm yields superior BLER performance with a minimal number of ports. Furthermore, extensive simulation results indicated that the OFDM-FAMA could support a significant number of UTs with robust channel coding. 

\vspace{-2mm}
\bibliographystyle{IEEEtran}

\end{document}